\documentclass[epj]{svjour}
\usepackage{epsfig,graphics,amssymb,wasysym}
\usepackage{colordvi}\usepackage[usenames]{color}
\newcommand{\etal}{{\it et al.}}
\newcommand{\edit}[1]{#1}
\begin{document}
\title{The quest for light multineutron systems}
\author{F.~Miguel Marqu\'es\inst{1} \and Jaume Carbonell\inst{2}
}                     


\institute{LPC Caen, Normandie Universit\'e, ENSICAEN, Universit\'e de Caen, CNRS/IN2P3, 14050 Caen, France 
\and Universit\'e Paris-Saclay, CNRS/IN2P3, IJCLab, 91405 Orsay, France}
\date{Received: date / Revised version: date}

%
\abstract{
 The long history of the research concerning the possible existence of bound or resonant states in light multineutron systems, essentially $^3$n and $^4$n, is reviewed. Both the experimental and the theoretical points of view have been considered, with the aim of showing a clear picture of all the different detection and calculation techniques that have been used, with particular emphasis in the issues that have been found. Finally, some aspects of the present and future research in this field are discussed.
%
\PACS{
      {Trineutron, Tetraneutron, Multineutron, Few-Body Systems, Few-Nucleon Problem, {\it ab initio}}{}
     } 
} 
\maketitle

\today

\bigskip \setcounter{tocdepth}{2} \tableofcontents \bigskip

\section{Introduction} \label{intro}

 Understanding the structure of nuclei and how their properties emerge from the underlying forces between nucleons is a major goal of modern nuclear physics.
 Light nuclei have long played a fundamental role in this respect, and those exhibiting very asymmetric neutron-to-proton ratios 
 have proven to be particularly sensitive to details of the two- and few-body forces used in nuclear models.
 Therefore, the question about the existence of multineutrons, the most extreme combinations one can find, raises many experimental and theoretical challenges.
 The discovery of such neutral systems as bound or resonant states would have far-reaching implications for many facets of nuclear physics, \edit{from the nature of the force itself up to the way it builds nuclei, as we will see in Sec.~\ref{Sec_Th}, and also for the modeling of neutron stars (see for example Ref.~\cite{4n_stars_USAL_EJPA155_2019})}.

 The quest for neutral nuclei 
 may be traced back to the early 1960s \cite{Ogloblin89}. 
 However, the last two decades have witnessed a renewed and enhanced interest in studying light neutron systems.
 This is essentially due to two experimental results claiming the possible observation of bound or low-lying resonant tetraneutron states \cite{Marques02_4n_recoil,Kisamori16_4n_DCX}, which motivated new experiments but mostly new theoretical studies, leading to progress in the computation of the exact solutions of the few-nucleon system in the continuum
 (see for example Refs.~\cite{FEWNC1,FEWNC2,FEWNC3,FEWNC4,FEWNC5,FEWNC6}).

 The main difficulty in the study of multineutron states
is that there are no particle-stable substructures, unlike standard charged nuclei. 
 If the dineutron were bound, one could more comfortably inquire about the existence of $^3$n (as $n+^2$n) or $^4$n (as $^2$n$+^2$n), both experimentally and theoretically.
 Despite some recent speculations \cite{WG_2n_PRC85_2012} this seems however totally excluded, making the progress in this field particularly non trivial.
 This intrinsic difficulty explains the lack of strong experimental evidence concerning the existence of such states, and leads on the theoretical side to contradictory results for the very same systems, even when described using the same $nn$ interaction. 

 As the title says, we have restricted this review paper to small clusters of neutrons, basically $^3$n, $^4$n and $^6$n, the ones that can be realistically accessed in experiments and that can be theoretically tackled with exact {\it ab initio} methods.
 We have deliberately omitted all the interesting studies concerning nuclear matter and neutron droplets, involving a large number of particles, for they require a very different approach from the experimental as well as the theoretical points of view.
 The interested reader can find helpful reviews in Refs.~\cite{n_drops,Ndrops0,Ndrops1,Ndrops2,Ndrops3}.

 The paper is divided in two main sections, corresponding to the review of the experimental (Sec.~\ref{Sec_Ex}) and theoretical (Sec.~\ref{Sec_Th}) works.
 On the experimental side, we have tried to class the long series of very different experiments, across a wide range of techniques, into a few categories for clarity, and in doing so we have restricted ourselves to openly accessible works. 
 In the theoretical part we have focused our efforts mainly on the results obtained during the last twenty years, which also coincide with the main experimental signals. 
 We will close the paper with some conclusions and perspectives in the field.

\section{Multineutron experiments}\label{Sec_Ex}

\def\longarrow{\relbar\!\!\relbar\!\!\relbar\!\!\relbar\!\!\relbar\!\!\relbar\!\!\rightarrow}
\def\hof{\hskip5.5ex}\def\inter{\\[-1.55ex]\hof$\Downarrow$\\[-.55ex]\hof$\Uparrow$\\[-1.6ex]}
\newcommand{\scheme}[3]{\begin{tabular}{l}#1\inter#2\\[0mm]\multicolumn{1}{c}{\normalsize#3}\end{tabular}}

\begin{figure}[t] \begin{center}
  \psfig{file=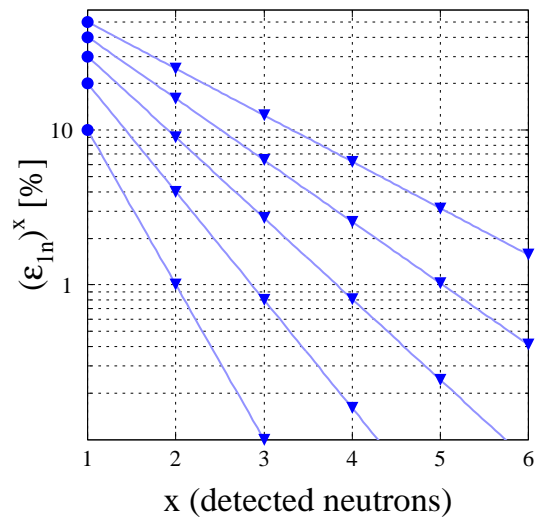,width=7cm}
 \end{center} \caption{The neutron efficiency to the power of the number of detected neutrons, as a function of the latter. From bottom to top, the lines/symbols correspond to neutron efficiencies $\varepsilon_{1n}$ of 10, 20, 30, 40 and 50\%, respectively. The vertical scale goes from 1\permil\ to 60\%.
 In practice, due to cross-talk effects the multineutron ($x>1$) efficiency will be even lower, $\varepsilon_{xn}<(\varepsilon_{1n})^x$.} \label{f:efficiency}       
\end{figure}

 Detecting neutral particles represents an experimental challenge in general. Charged particles crossing a detector material interact with the atomic electron clouds, leading to detection efficiencies of 100\%. Neutral particles, however, must interact with a nucleus, a much less probable process, leading to typical detection efficiencies (for similarly sized detectors) of few per cent.
 Moreover, the detection of $x$ neutrons becomes exponentially harder, since the efficiency will decrease roughly as $\varepsilon_{xn}\approx(\varepsilon_{1n})^{\,x}$, as shown in Fig.~\ref{f:efficiency}. To make things worse, the latter is just an upper limit due to cross-talk effects \cite{cross-talk} (a neutron may interact several times), that require the application of rejection algorithms with subsequent losses of total efficiency. 

 Even if one obviates the detection problem, the fact that neutrons are difficult to guide and unstable themselves does not allow us to build a system of several neutrons out of its components.
 Therefore, all the experimental approaches must face the challenge of overcoming those two issues: how to build multineutrons and then how to detect them. In reviewing all the existing works since the 1960s we have followed the chronology when possible, but at the same time we have tried to organize them in subsections according to the probe they used.
 
 Paradoxically, even if the aim of these experiments is the `observation' of systems of several neutrons, almost all of them have in common the absence of any neutron detection, due to the aforementioned very low $\varepsilon_{xn}$ efficiencies. 
 We have identified basically two main categories of experiments according to their principle:

{\Large \begin{center}
 \scheme{\textcircled{\it{a}}$\longarrow\,$\textcircled{\it{b}}}
        {\textcircled{\it{c}}$\longarrow\!(^A\mbox{n})$}{\it missing mass} \hskip2.ex
 \scheme{\textcircled{\it{a}}$\longarrow\,$\textcircled{\it{b}}}
        {\hskip-1.8ex$(^A\mbox{n})\!\longarrow\!(X)$}{\it two step}
 \end{center}}

\noindent In both of them one must detect in the final state only one charged particle, \textcircled{\it{b}}, either directly or indirectly.

 On the left panel, representing the missing-mass category, the multineutron is sought to be formed in a two-body collision \textcircled{\it{a}}+\textcircled{\it{c}} leading to a two-body final state \textcircled{\it{b}}+$^A$n.
 Only in that case, the detection of \textcircled{\it{b}} can sign the population of states in the missing multineutron system through the unique constraints of two-body kinematics. The main advantages ($+$) and issues ($-$) of the missing-mass technique are:
 \begin{description}
 \item$+$ It only requires the detection of one charged particle.
 \item$+$ The multineutron mass number $A$ is well defined.
 \item$+$ Both bound and resonant $^A$n states can be probed.
 \item$-$ The cross-sections that bring all the protons into \textcircled{\it{b}} without breaking it are generally (extremely) low.
 \item$-$ Any beam or target contaminant different from \textcircled{\it{a}} or \textcircled{\it{c}} would lead to a missing partner(s) of \textcircled{\it{b}} that is not a multineutron.
 \end{description}
 
 On the right panel above, representing the two-step category, a bound multineutron is supposed to be produced in a previous reaction, and then one seeks to sign its subsequent interaction with a nucleus \textcircled{\it{a}} that induces a transformation into \textcircled{\it{b}}. The advantages and issues of the two-step category are:
 \begin{description}
 \item$+$ It only requires the detection of one charged particle.
 \item$-$ Only bound $^A$n states can lead to this second reaction.
 \item$-$ There is no sensitivity to the multineutron energy.
 \item$-$ Only a lower limit of the mass number $A$ can be inferred from the transformation into \textcircled{\it{b}}.
 \item$-$ A contaminant different from \textcircled{\it{a}} could lead to \textcircled{\it{b}} without requiring a multineutron.
 \item$-$ The generally uncontrolled previous reaction may produce a huge background of many particles, that could be eventually responsible for the production of \textcircled{\it{b}}.
 \end{description}
 
 Concerning the mass number of the multineutron candidate systems, although $^3$n is the simplest one, the pairing effects observed along the neutron dripline suggest that $^{4,6,8}$n could more likely exhibit bound or resonant states.
 However, the availability of beams and targets, and the significance of the cross-sections involved, have mostly limited the search to $^{3,4}$n.
 This is particularly true in the missing-mass technique, since the charged partners define the multineutron mass number and $A>4$ would require too unbalanced partners, and/or negligible cross-sections.
 On the other hand, the ambiguity in mass number of two-step reactions has opened the (uncertain) way to the search for heavier multineutrons, as we will see.

 Therefore, due to accessibility and pairing considerations, the tetraneutron has been the most natural aim of these searches. 
 Concerning its eventual binding energy, an upper limit is provided by the minimum of the experimental $4n$ separation energy ($S_{4n}$) of bound nuclei, since a higher tetraneutron binding would make them unbound. For most of the quest this value was $S_{4n}(^8$He$)=3.1$~MeV, but recent mass measurements have decreased it to $S_{4n}(^{19}$B$)=1.5(4)$~MeV \cite{AME2016}.
 However, if the tetraneutron were bound by more than 1~MeV, $\alpha+^4$n would be the first particle threshold in $^8$He. As the breakup of $^8$He is dominated by the $^6$He channel \cite{Warner_6He}, the tetraneutron, if bound, should be so by less than 1~MeV.  

 Before starting the detailed review of the experiments, note that we will only address the results that have been published in accessible international journals. Moreover, we will not cover experiments that tried to probe the existence of multineutrons indirectly through the search of analog states in neighboring nuclei, like $^{3,4}$H and $^{3,4}$He.

\subsection{The pion probe} \label{sec:Pions}

 One of the cleanest probes is the double charge exchange (DCX) reaction $(\pi^-,\pi^+)$, which falls within the missing-mass category. The incoming negative pion becomes positive, changing two protons in the target into neutrons. By measuring the momenta of the pions on a $^{3,4}$He target, one can measure the missing mass of the $^{3,4}$n system. Obviously, since no other helium isotopes can be used as targets, no other multineutrons are accessible.
 
\begin{figure}[ht] \begin{center}
  \psfig{file=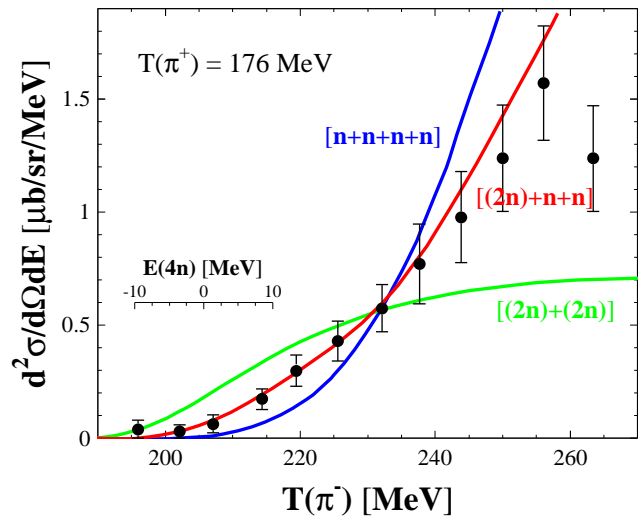,width=8.5cm}
 \end{center} \caption{Experimental results for the reaction $^4$He$(\pi^-,\pi^+)4n$ at $\theta=0^\circ$. The curves represent the phase space for the final states $n+n+n+n$, $(2n)+n+n$, and $(2n)+(2n)$, normalized at 232~MeV and including the experimental conditions. The small axis represents the energy of the neutrons around the $4n$ threshold. Adapted from Ref.~\cite{Gilly65_4n_pi4He}.} \label{f:Gilly}       
\end{figure}
 
 To our knowledge, this axis started in 1965 when Gilly \etal\ searched for the $^4$He$(\pi^-,\pi^+)^4$n reaction \cite{Gilly65_4n_pi4He}. They found no evidence of the tetraneutron, that would manifest itself through peaks in the $\pi^+$ spectrum due to two-body kinematics.
 The spectrum could be explained by introducing a final-state interaction (FSI) between two of the neutrons, as can be seen in Fig.~\ref{f:Gilly} (red curve).

 In 1970 Sperinde \etal\ used a $^3$He target and searched for the $^3$He$(\pi^-,\pi^+)^3$n reaction \cite{Sperinde70_3n_pi3He}. No evidence for a bound trineutron was found, but an enhancement at low $3n$ energies was observed. The same enhancement was observed in 1974 \cite{Sperinde74_3n_pi3He}, but it was finally explained by the FSI between two of the neutrons or between all of them in a two-step DCX \cite{Jibuti85_34n_pi34He}.
 
 In 1976 Bistirlich \etal\ used a variant of this technique by studying the single charge exchange (SCX) on hydrogen, $^3$H$(\pi^-,\gamma)3n$ \cite{Bistirlich76_3n_pi3H}. After the subtraction of the different contaminants present in the tritium cell, no evidence for a trineutron was found in the $\gamma$ spectrum. In 1980, Miller \etal\ undertook the same study with better statistics and resolution \cite{Miller80_3n_pi3H}, and confirmed that no trineutron was needed to describe the observed spectrum, only the $2n$ FSI was required.
 
\begin{figure}[ht] \begin{center}
  \psfig{file=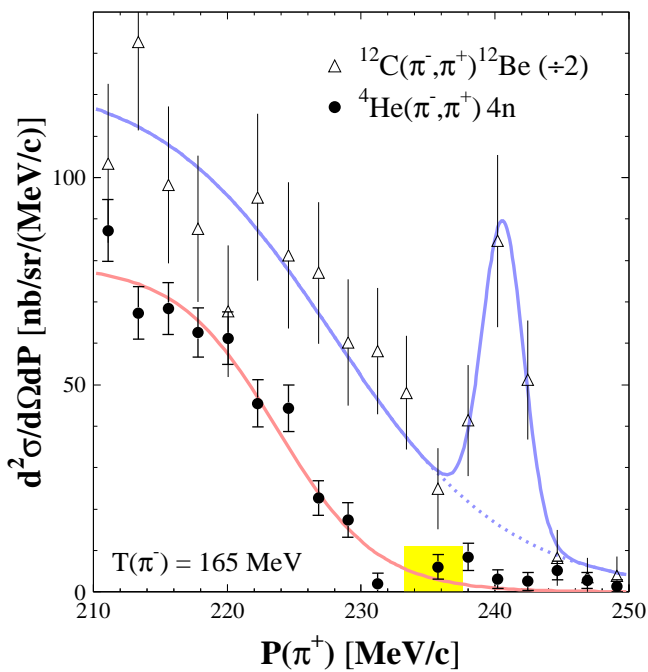,width=8.5cm}
 \end{center} \caption{Experimental results for the reactions $^4$He$(\pi^-,\pi^+)4n$ at $\theta=0^\circ$ (circles) and $^{12}$C$(\pi^-,\pi^+)^{12}$Be at $\theta=8^\circ$ (triangles, divided by 2). The curves are fits to guide the eye, with a Woods-Saxon distribution only (red) plus an additional Gaussian function (blue). The peak in the $^{12}$C channel corresponds to the formation of the $^{12}$Be ground and two first excited states, and the range in yellow in the $^4$He channel to the region expected for a bound tetraneutron. Adapted from Ref.~\cite{Ungar84_4n_pi4He}.} \label{f:Ungar}       
\end{figure}

 In 1984 Ungar \etal\ revisited the tetraneutron with the $^4$He$(\pi^-,\pi^+)^4$n reaction \cite{Ungar84_4n_pi4He}.
 As a test, they also measured the $^{12}$C$(\pi^-,\pi^+)^{12}$Be reaction, in which the formation of states in $^{12}$Be appeared clearly as a peak in the $\pi^+$ spectrum (Fig.~\ref{f:Ungar}, triangles).
 In the $4n$ channel, however, no apparent peak was observed (Fig.~\ref{f:Ungar}, circles).
 A few events laid in the expected region of bound $^4$n, although at a rate similar to the events beyond that region, which are kinematically forbidden. The latter background was due to imperfect rejection of the events in which the $\pi^+$ decayed inside the spectrometer.
 In search of a possible resonant behavior, the spectrum below the $4n$ threshold was found to exhibit a broad enhancement with respect to four-neutron phase space, but it was consistent with the FSI of two neutron pairs \cite{Ungar84_4n_pi4He}.
 
 In 1986, Stetz \etal\ extended the search to the trineutron with the reactions $^{3,4}$He$(\pi^-,\pi^+)^{3,4}$n \cite{Stetz86_34n_pi34He}, at several energies and angles, but again no evidence of trineutron or tetraneutron states was found.
 
 In 1989 Gorringe \etal\ repeated Ungar's experiment at lower energy, to reduce the continuum DCX contribution, but at a higher angle, corresponding to a higher momentum transfer to the neutrons \cite{Gorringe89_4n_pi4He}. 
 After background subtraction, 6~counts remained in the possible bound $^4$n window, but those were consistent with the estimated continuum contribution. Only an upper limit of the production cross-section could be deduced.

 In 1997, a systematic study of the $^3$He$(\pi^-,\pi^+)3n$ reaction by Yuly \etal\ \cite{Yuly97_3n_pi3He} at 120--240~MeV did not show evidence for a trineutron, and indicated that the DCX process proceeds as a sequential SCX one.
 In 1999, Gr\"ater \etal\ studied the same reaction at lower energies, 65--120~MeV, and found no evidence either \cite{Grater99_3n_pi3He}.
 
 For the sake of completeness, let us mention two special, non-conventional uses of the negative-pion probe.
 Earlier, in 1979, Chultem \etal\ had proposed an original use of a $\pi^-$ beam in a two-step process \cite{Chultem79_4n_pi208Pb}. They hoped to produce a bound tetraneutron in the $^{208}$Pb$(\pi^-,\pi^+)^4$n reaction, by DCX on an $\alpha$ cluster inside lead, and then measure the tetraneutron absorption by another lead nucleus, leading to $^{212}$Pb. However, they found no $\alpha$ particles from its decay chain into $^{212}$Bi and then $^{212}$Po.
 Finally, in 1991 Gornov \etal\ studied the $3n$ missing mass in the reactions $^9$Be$(\pi^-,t\,^3$He$)$ and $^9$Be$(\pi^-,d\,^4$He$)$ \cite{Gornov91_3n_pi9Be}. The missing-mass spectrum was tentatively described using a very broad trineutron resonance at 3~MeV, but the very limited resolution did not allow to draw firm conclusions. 

 In summary, after more than 30~years of DCX experiments with pion beams, only upper limits following the non-observation of multineutrons have been set, and the technique is not being presently used.
 However, the DCX technique has been recently revisited by Kisamori \etal\ using exotic nuclei \cite{Kisamori16_4n_DCX}, as we will see in Sec.~\ref{sec:RIKEN16}.
 
\subsection{Fission and other activation probes}

 The DCX pion reactions are a very clean and powerful probe, although the cross-sections involved are extremely low. From the start of the multineutron program a second avenue was taken, exploiting the two-step principle described previously.
 In the first step, the multineutron is supposed to be produced in a high-flux reaction, not necessarily well-characterized, like spallation or induced fission. Assuming that a bound multineutron was thus produced, together with many other particles, we may use it in a secondary two-body reaction that transforms a given sample in a unique way.
 Therefore, we need to demonstrate that the sample was transformed, but above all that it could not be transformed in any other alternative way.

 Already in 1963, Schiffer \etal\ irradiated samples of nitrogen and aluminum in a nuclear reactor, searching for the reactions $^{14}$N$(^4$n$,n)^{17}$N and $^{27}$Al$(^4$n$,t)^{28}$Mg \cite{Schiffer63_4n_nU}. If bound tetraneutrons had been produced in the fission of the uranium fuel, they might have observed the $\beta$-delayed neutron decay of $^{17}$N and the $\gamma$-rays from $^{28}$Mg and its daughter $^{28}$Al. However, none of them were significantly observed above the background.
 
 In 1965 Cierjacks \etal\ carried out a similar experiment, with samples of nitrogen, oxygen, magnesium and other heavier elements surrounding a uranium target bombarded with deuterons \cite{Cierjacks65_4n_d238U}. For the three lighter samples, they searched for bound tetraneutrons emitted in uranium fission through the reactions $^{14}$N$(^4$n$,n)^{17}$N, $^{16}$O$(^4$n$,t)^{17}$N and $^{26}$Mg$(^4$n$,2n)^{28}$Mg. However, they were not able to observe the corresponding $\beta$-delayed neutrons or $\gamma$-rays above the background.

\edit{In 1968 Fujikawa \etal\ searched for bound trineutrons in the reaction $n+^7$Li, that would transform a barium sample through the secondary reaction $^{138}$Ba$(^3$n$,n)^{140}$Ba \cite{Fujikawa68_3n_nLi}. Barium was in the form of a high-purity BaCO$_3$ powder, and the potential production of $^{140}$Ba would have been identified through the $\gamma$-rays of its daughter, $^{140}$La. After chemical separation of lanthanum from the powder, no $\gamma$-rays were observed, and an upper limit of the cross-section was provided.}

\begin{figure}[ht] \begin{center}
  \psfig{file=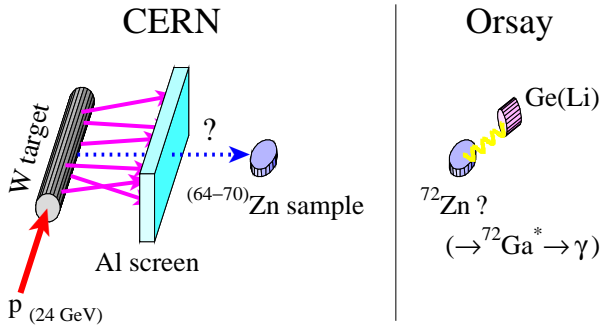,width=8.5cm}
 \end{center} \caption{Schematic view of D\'etraz's experiment \cite{Detraz77_68n_pW}. In the collision of protons and tungsten a neutral cluster was supposed to be produced, go through a screen, and induce $(^A$n$,xn)$ reactions on natural zinc. The new $^{72}$Zn isotope was detected through its decay into $^{72}$Ga$^*$ with a germanium detector.} \label{f:Detraz}       
\end{figure}

 In 1977 the multineutron search seemed to come to a happy end at CERN.
 D\'etraz obtained the first positive result after exposing a natural zinc sample to a tungsten block irradiated with a proton beam of 24~GeV \cite{Detraz77_68n_pW} (Fig.~\ref{f:Detraz}). Block and sample were separated by an aluminum screen, supposed to be thick enough to stop any charged particle that could have been produced, and the reaction that was sought was $^{64,66,67,68,70}$Zn$(^A$n$,xn)^{72}$Zn, i.e.\ the transformation of a natural zinc isotope into $^{72}$Zn through the interaction with a neutral cluster.
 The relatively long half-life of $^{72}$Zn (46.5~h) and its daughter $^{72}$Ga (14.1~h), and the several $\gamma$-rays emitted by the latter, made it possible to remove the sample from the high-activity area and perform a clean measurement of those $\gamma$-rays.

 After several-day exposure, the 100~g of natural zinc were sent by air freight to Orsay, where a Ge(Li) detector revealed the unambiguous production of $^{72}$Zn through five prominent $\gamma$-rays of its daughter $^{72}$Ga.
 This observation required up to four chemical separations of gallium, due to the high activity induced in the sample by other beam-induced reactions. 
 The production of $^{72}$Zn could have been the result of any ($A>1$) bound multineutron, but since the previous searches for $A=2$--4 had failed, D\'etraz concluded that most likely bound hexa- or octaneutrons ($A=6,8$) had been produced \cite{Detraz77_68n_pW}.
 
 Only a few months later, Turkevich \etal\ tried to confirm D\'etraz's exciting results using a similar principle \cite{Turkevich77_6n_pU}. As the first step, they irradiated uranium with 700~MeV protons, and exposed a lead sample behind an aluminum screen, again supposed to stop any charged particle. 
 For the second step, they used the one that was described in the previous section for Chultem \cite{Chultem79_4n_pi208Pb}, the search for the $^{208}$Pb$(^A$n$,xn)^{212}$Pb reaction through the $\alpha$ particles from its decay into $^{212}$Bi and $^{212}$Po.
 No trace of ``polyneutrons'' was found, shedding doubt on D\'etraz's results.
 
 In 1980, de Boer \etal\ followed suit. Since Turkevich concluded that their results would have not been sensitive to multineutrons with $A<6$ \cite{Turkevich77_6n_pU}, a possible solution to D\'etraz's puzzle was that he had observed only $^4$n, and not $^{6,8}$n as he had suggested \cite{Detraz77_68n_pW}.
 Therefore, de Boer irradiated a block of tellurium with $^3$He ions and searched for the interaction of bound tetraneutrons in the same block through the $^{130}$Te$(^4$n$,2n)^{132}$Te reaction \cite{deBoer80_46n_3He130Te}.
 No $\gamma$-rays from the decay of $^{132}$Te were observed, and de Boer concluded, in agreement with D\' etraz, that the latter had in fact underestimated the transmission of tritons through his aluminum shield and that the reaction observed had been $^{\rm(nat)}$Zn$(t,p)^{72}$Zn, without any multineutron \cite{deBoer80_46n_3He130Te}.

 This tree-year period of excitement ended up with some frustration, but the multineutron quest went on. However, following these results, the two-step or activation probe was abandoned for more than thirty years in favor of the cleaner missing-mass experiments.

 Anyhow, in 2012 Novatsky \etal\ revisited the technique by inducing uranium fission with 62~MeV $\alpha$ particles, and then searched for the interaction of bound multineutrons within strontium and aluminum samples shielded by thin Kapton and beryllium layers \cite{Novatsky12_6n_a238U,Novatsky13_6n_a238U}.
 The authors claimed they had observed the reaction $^{88}$Sr$(^A$n$,xn)^{92}$Sr, through $\gamma$-rays from the daughter $^{92}$Y \cite{Novatsky12_6n_a238U}, and the reaction $^{27}$Al$(^A$n$,p\,xn)^{28}$Mg, through $\gamma$-rays from the decay into $^{28}$Al and $^{28}$Si \cite{Novatsky13_6n_a238U}.
 They concluded that, due to the lack of evidence for a bound tetraneutron, their results should ``certainly'' correspond to heavier clusters ($A\geqslant6$).
 However, the experimental hall exhibited strong (and diverse) activity, 
 the number of reaction channels following fission is extremely high, and it could be that the screens that were used were not thick enough to shield all the many charged particles produced, as it had been the case for D\'etraz in 1977.
 
 To conclude, the so-called activation probe appeared as a powerful tool from the very first years, lead to some hope about bound multineutrons in the late 70s, but after the refutation of the claim it was left out and is not being followed nowadays in any large-scale facility.

\subsection{The multinucleon-transfer probe} \label{sec:Transfer}

 A third research axis was also opened in the early years. Like the pion DCX, it relied on the cleaner missing-mass principle. However, instead of changing two protons into two neutrons, which limited the search to $^{3,4}$He targets, the multineutron system was probed in a transfer reaction between two stable nuclei, increasing the possible combinations.
 But as in the DCX reaction, the cross-sections were expected to be very low, since one needed to: transfer all the protons away from one nucleus; sometimes bring some neutrons back in the opposite sense; and in any event lead exclusively to a two-body final state.
 
 In 1965 Ajda\v{c}i\'c \etal\ started to use this probe with a simple $(n,p)$ transfer reaction on the triton \cite{Ajdacic65_3n_n3H}. With a 14~MeV neutron beam they detected protons from the $^3$H$(n,p)3n$ reaction. Some events were observed at a missing mass of about 1~MeV below the $3n$ threshold, the first candidates of a (quite) bound trineutron. However, knowing that the very first experiments had failed to find a bound tetraneutron, more likely to exist due to pairing, they concluded that their result was ``highly improbable''.
 One year later, Thornton \etal\ repeated the same experiment with 21~MeV neutrons and better resolution \cite{Thornton66_3n_n3H}, and found no evidence for a bound trineutron.
 
 In 1968 Ohlsen \etal\ used a triton beam and searched already for a more complex transfer reaction, $^3$H$(t,^3$He$)3n$ \cite{Ohlsen68_3n_t3H}. The missing mass reconstructed from $^3$He lead to a deviation from four-body phase space, only at forward angles, that could be consistent with a low-energy trineutron resonance. They were not able to exclude, however, an effect from the reaction mechanism itself.

 In 1974 Cerny \etal\ started to use heavier nuclei and searched for the tri- and tetraneutron in the reactions $^7$Li$(^7$Li$,^{11}$C$)3n$ and $^7$Li$(^7$Li$,^{10}$C$)4n$ \cite{Cerny74_43n_transfer}. 
 Concerning the trineutron, the intense $^{11}$C channel led to a very high statistics spectrum. Unlike the $^7$Li$(^7$Li$,^{11}$B$)^3$H channel, that exhibited several structures, the $3n$ missing-mass spectrum could be well described by four-body phase space, plus some small peaks from known target contaminants (that lead to $^{11}$C partners different from $3n$). 
 
\begin{figure}[ht] \begin{center}
  \psfig{file=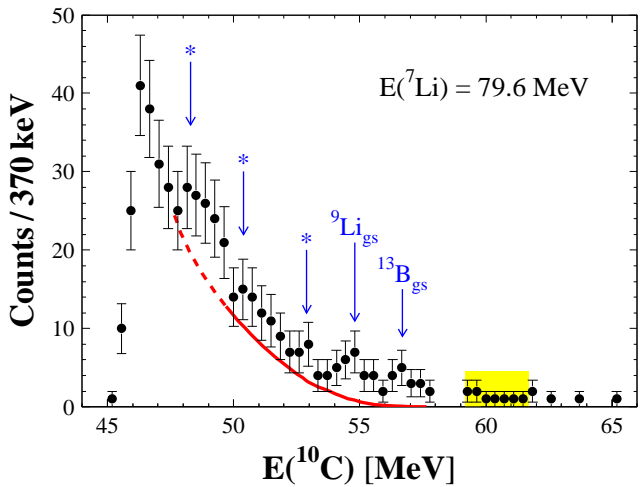,width=8.5cm}
 \end{center} \caption{Energy spectrum of $^{10}$C from the $^7$Li$(^7$Li$,^{10}$C$)4n$ reaction at $\theta=7.4^\circ$. Known contaminant reactions are indicated either explicitly or with an asterisk. The red curve corresponds to five-body phase space, and the range in yellow to the region expected for a bound tetraneutron. Adapted from Ref.~\cite{Cerny74_43n_transfer}.} \label{f:Cerny}       
\end{figure}

 In the tetraneutron channel, however, the low $^{10}$C production led to a poor separation from the tail of the much stronger $^{11}$C distribution. The resulting $4n$ missing-mass spectrum could be described by five-body phase space plus the known contaminants, as can be seen in Fig.~\ref{f:Cerny}. Although some events are visible in the possible region for bound $^4$n, the signal was not significant with respect to the background level.
 They concluded that the purity was still much worse than in the DCX experiments, but that by using different beam-target combinations the way had been opened to heavier multineutrons.

 In 1988 Belozyorov \etal\ improved on the main issues of Cerny's work, the target purity and fragment identification, and searched for the trineutron in the $^7$Li$(^{11}$B$,^{15}$O$)3n$ reaction and for the tetraneutron in the $^7$Li$(^{11}$B$,^{14}$O$)4n$, $^7$Li$(^9$Be$,^{12}$N$)4n$ and $^9$Be$(^9$Be$,^{14}$O$)4n$ reactions \cite{Belozyorov88_43n_transfer}.
 In all the reactions on $^7$Li, the missing-mass spectra above the $3n$ and $4n$ thresholds were well described by the corresponding four- and five-body phase space, showing no evidence for multineutron resonances. 
 Although the two reactions leading to $^{14}$O showed a few events below the $4n$ threshold, they were consistent with the background due to pulse pileup or to beryllium target impurities \cite{Belozyorov88_43n_transfer}.
 
 In 1995 Bohlen \etal\ used a $^{14}$C beam in order to probe very neutron-rich missing masses, among them the trineutron in the reaction $^2$H$(^{14}$C$,^{13}$N$)3n$ on a CD$_2$ target \cite{Bohlen95_3n_transfer}. The missing mass was fully described below the threshold by the carbon contribution and above it by the decay of a broad $^{15}$N resonance.
 Finally, in 2005 Aleksandrov \etal\ repeated Cerny's experiment, with similar beam energy and target, and obtained the same negative results for both the tri- and tetraneutron \cite{Aleksandrov05_43n_transfer}.
 
 This technique appeared as a good compromise between pion DCX and activation. It could access heavy multineutrons from a variety of beam-target combinations, and the potential signals were supposed to be unambiguous. However, the absence in practice of clear signals led towards a need for higher purities, and the technique has been put aside for the last fifteen years.

\subsection{The GANIL 2002 result}

 The experiments performed in the XX century used mainly stable beams and targets. The beams could thus be very intense, but building a neutral system from balanced combinations of protons and neutrons required reactions with very low cross-sections.
 Moreover, the potential multineutron signal often shared parts of the spectra with background from contaminant species, and due to the low cross-sections used the background contributions became too important for a signal to be clearly established.

 In 2002 Marqu\'es \etal\ proposed at the GANIL facility a new technique that could solve those issues \cite{Marques02_4n_recoil}. With the advent of radioactive secondary beams, the possible preformation of multineutrons inside very neutron-rich nuclei was considered, similar to the preformation of $\alpha$ particles in the process of $\alpha$ decay.
 Within this scenario, the until then complex formation step of multineutrons could be reduced to the breakup of one of those nuclei,
 with an increase in cross-section of several orders of magnitude (mb, compared to the nb or pb of the previous probes) due to the weak binding of these clusters.
  
 The radioactive beam was $^{14}$Be, in which the $^{10}$Be$+4n$ threshold is at only 5~MeV. Following the breakup on a carbon target at 35~MeV/N, the detection of the $^{10}$Be fragment provided a clean signature of the channel.
 For the detection of a potentially liberated tetraneutron cluster, the principle was similar to that used by Chadwick in the discovery of the neutron \cite{Chadwick32}: deduce the mass of the neutral particle from the recoil induced by elastic scattering on charged particles.
 The recoil energy $E_p$ of a proton in the organic scintillator detector is related to the energy per nucleon $E_n$ of the incoming $^A$n cluster, obtained from its time of flight: $(E_p/E_n)\leqslant4A^2/(A+1)^2$.
 Since the dineutron is unbound, and the trineutron should also be due to pairing, the measurement of proton recoils over the incoming neutron energy of 1.5--2.5 could only be attributed to a bound tetraneutron.
 
\begin{figure}[t] \begin{center}
  \psfig{file=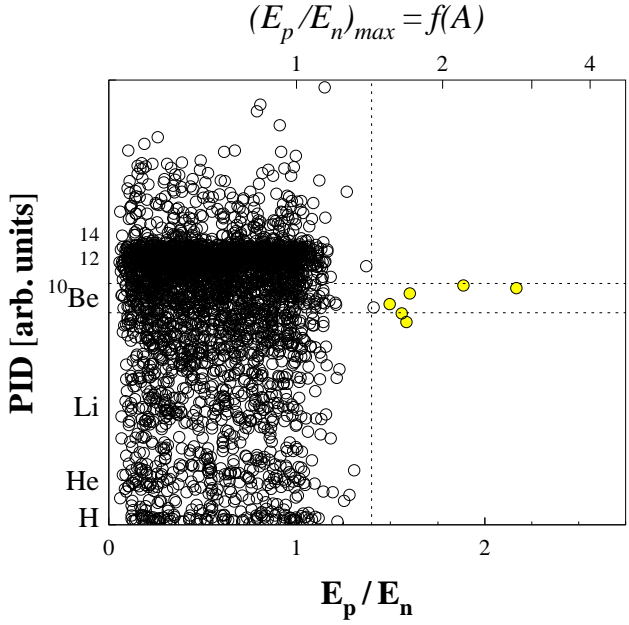,width=8.5cm}
 \end{center} \caption{Scatter plot of the particle identification parameter PID vs the proton recoil in the neutron detector (normalized to the neutron energy) for the reaction $(^{14}$Be$,X+n)$. The dotted lines show the region centered on the $^{10}$Be peak and with $E_p/E_n>1.4$, and the 6~events in yellow are candidates to the formation of a bound tetraneutron. The scale on the upper axis shows the maximum proton recoil as a function of the multineutron mass number. Adapted from Ref.~\cite{Marques02_4n_recoil}.} \label{f:GANIL}       
\end{figure}

 The method was also applied to data from $^{11}$Li and $^{15}$B beams, but only in the case of $^{14}$Be some events were observed with characteristics consistent with the production and detection of a multineutron cluster. The 6~events are shown in Fig.~\ref{f:GANIL} in yellow, with proton recoils 1.4--2.2 times higher than those expected for individual neutrons, and appear all in coincidence with the detection of a $^{10}$Be fragment (note the absence of events in the most abundant $^{12}$Be channel). 
 Special care was taken to estimate the effects of pileup, i.e.\ the detection of several neutrons in the same detector, and it was found that it could at most account for some 10\% of the observed signal. The conclusion for the most probable scenario was the formation of a bound tetraneutron in coincidence with $^{10}$Be \cite{Marques02_4n_recoil}.
 
 As we will see in Sec.~\ref{Sec_Th}, this result triggered several theoretical calculations that could not explain the possible binding of the tetraneutron. Moreover, another work questioned the probe itself \cite{Sherrill_Bertulani}, arguing that a weakly bound tetraneutron would rather undergo breakup than elastic scattering.
 Marqu\'es \etal\ addressed both issues \cite{Marques05_4n_recoil}, finding that the signal observed 
 could be generated also by a low-energy tetraneutron resonance ($E\lesssim2$~MeV)
 through an enhancement of pileup,
 or by the breakup of a bound tetraneutron in the scintillator followed by the detection of some of the neutrons.
 
 Interestingly, the latter scenario had been already suggested by Brill \etal\ in 1964 as a potential probe of bound multineutrons, but no abnormal signals were observed at that time in the bombardment of a $^{48}$Ca target with $^{12}$C and $^3$He beams \cite{Brill64_46n_recoil}.
 \edit{On the other hand, another experiment in 1971 by Koral \etal\ had already tried to detect bound trineutrons from abnormal induced recoils \cite{Koral71_3n_nLi}. Following the bombardment of a lithium target with neutrons similar to the one in Ref.~\cite{Fujikawa68_3n_nLi}, they searched abnormal recoils of $^4$He ions by comparing the time of flight and the pulse height signals in a helium scintillator, the same principle used at GANIL, but no evidence was found.}
 
 The decrease in beam intensities at GANIL and the aging of the neutron detector used did not allow an unambiguous confirmation of this result.
 Other experiments at GANIL using the missing-mass technique were not able to find positive signals either. After some theoretical works in the early 2000s, 
 the field became quiet.

\subsection{The RIKEN 2016 result} \label{sec:RIKEN16}

 In 2016, Kisamori \etal\ proposed a new probe at RIKEN: $^4$He$(^8$He$,^8$Be$)^4$n, a DCX reaction using exotic nuclei \cite{Kisamori16_4n_DCX}.
 Sending a very intense $^8$He beam at 186~MeV/N onto a liquid $^4$He target, the exit channel was selected through the detection in an spectrometer of the two $\alpha$ particles from the decay in flight of $^8$Be.
 The large $Q$ value of the $(^8$He$,^8$Be$)$ reaction almost compensated the binding energy of the $\alpha$ particle and allowed for the formation of a $4n$ system with small momentum transfer.
 The authors expected in this way to enhance the odds of a weakly interacting tetraneutron in the final state, that would remain in the target area (and would not be directly detected).

\begin{figure}[t] \begin{center}
  \psfig{file=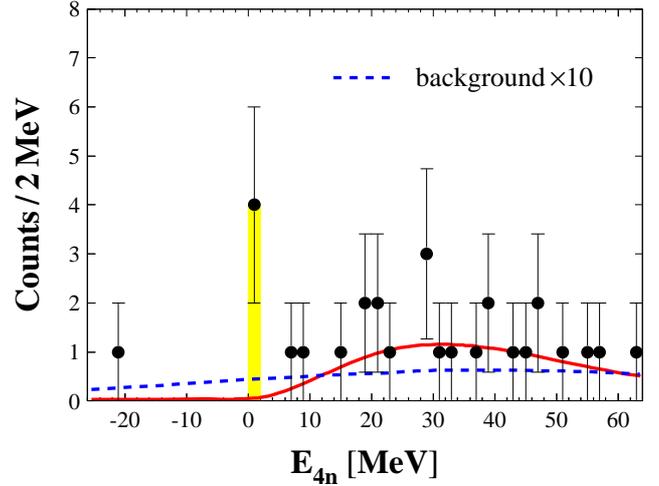,width=8.5cm}
 \end{center} \caption{Missing-mass spectrum of the $^4$He$(^8$He$,^8$Be$)4n$ reaction. The solid (red) curve represents the sum of the direct decay of correlated $2n$ pairs plus the estimated background. The dashed (blue) curve represents only the latter, multiplied by a factor of 10 in order to make it visible.
 The 4~events at threshold are highlighted in yellow.
 Adapted from Ref.~\cite{Kisamori16_4n_DCX}.} \label{f:Kisamori}
\end{figure}

 The $4n$ missing-mass spectrum (Fig.~\ref{f:Kisamori}) showed 4~events very close to threshold.
 The relative energy and angle between the two $\alpha$ particles were consistent with the formation of the $^8$Be ground state. The background was estimated from the probability of having two $^8$He beam particles from the same bunch breaking up, due to the high beam intensity, and leading to the detection of two independent $\alpha$ particles at small angle that could mimic a $^8$Be decay. It was found to be uniformly spread over the whole range of missing mass (blue curve in Fig.~\ref{f:Kisamori}), with an integrated value of about 2~events \cite{Kisamori16_4n_DCX}.

 The red curve in Fig.~\ref{f:Kisamori} corresponds to a calculation of the direct decay of the $4n$ final state within a wave packet similar to the initial $^4$He, including the interaction between neutrons and between neutron pairs \cite{Kisamori16_4n_DCX}. It includes also the estimated background described above, not visible at the scale of the figure. This curve, without the hypothesis of a tetraneutron resonance state, clearly cannot explain the events observed around the $4n$ threshold.
 Note that there was only 1~event in the kinematically forbidden region (at about $-20$~MeV in Fig.~\ref{f:Kisamori}), an additional indication of the low background.

 The 4~events in the region $0<E_{4n}<2$~MeV were found to be consistent with the formation of a tetraneutron resonance at $E(^4$n$)=0.8\pm1.3$~MeV, with a width $\Gamma<2.6$~MeV, taking into account both the statistic and systematic errors (the latter were dominant due to the missing-mass calibration procedure). It should be noted that, because of the missing-mass uncertainty, this result is also consistent with the formation of a bound tetraneutron. The cross-section corresponding to the candidate events was estimated at about 4~nb,
 compatible with a simple estimation of the two-step process assuming the
Gamow-Teller and the spin-dipole transition for the projectile
and the target system, respectively \cite{Kisamori16_4n_DCX}.

\subsection{The neutron probe}

 As mentioned when introducing the experimental techniques, paradoxically almost none of the many experiments already reviewed has tried to detect the neutrons. 
 With the recent improvements in beam intensities and neutron detection efficiencies in the world leading facilities, trying to detect the neutrons from the decay of a multineutron system seems a logical next step, which in addition would give access to eventual neutron correlations in their decay:
 
{\Large \begin{center} \scheme{\textcircled{\it{a}}$\longarrow\!X$}
  {\textcircled{\it{c}}$\longarrow\!(^A\mbox{n})\!\longrightarrow A$\textcircled{$n$}}{\it neutron detection} \end{center}}

 For the sake of completeness, we note that in 2016 Bystritsky \etal\ claimed the first direct detection of multineutrons with $A=6,8$ \cite{Bystritsky16_6n_natU}. They used an array of 20 $^3$He counters, with a total estimated efficiency of $\varepsilon_{1n}\sim20\%$, surrounding a $^{238}$U sample, and searched for the neutrons that would stem from the multineutron cluster decay of uranium.
 Although they based their claim in the detection of few events with 5~neutron hits, they also concluded that the confirmation would require an increase of the statistics, and a reduction of the cosmic and natural backgrounds, by at least one order of magnitude \cite{Bystritsky16_6n_natU}.
 
 The Radioactive Isotope Beam Factory (RIBF) of the RIKEN Nishina Center provides nowadays the highest intensities of light neutron-rich beams, together with high neutron-detection efficiencies.
 Since the 2016 result, two experiments have been undertaken at RIKEN aiming at the detection, for the first time, of all the neutrons emitted in a multineutron ($x>2$) decay.
 They benefited both from the combined capabilities of the NEBULA \cite{NEBULA} and NeuLAND demonstrator \cite{NeuLAND} neutron detector arrays. With an average neutron efficiency of $\varepsilon_{1n}\sim45\%$, the granularity of the arrays allowed for an unprecedented four-neutron efficiency of $\varepsilon_{4n}\sim1$--2\% (see Fig.~\ref{f:efficiency}).

 The first experiment searched for the ground state of $^{28}$O, one of the grails of nuclear structure physics, using the reaction H$(^{29}$F$,^{24}$O$)4n$ \cite{NP_Kondo}. The moderate intensity of the very exotic $^{29}$F beam lead to some 100 complete $4n$ events. Although the analysis of the results is still in progress, it seems that the decay of the ground state of $^{28}$O would proceed through the narrow $^{26}$O ground state, i.e.\ it would be a sequential $2n$-$2n$ decay. As such, no information should be derived concerning multineutrons.
 
 The second experiment used a similar setup and technique, the proton removal from a beam and the detection of a fragment plus four neutrons, but this time in search of the $^7$H ground-state energy and its potential tetraneutron decay with the reaction H$(^8$He$,^3$H$)4n$ \cite{NP_Marques}. With respect to the previous experiment, the much higher intensity of the $^8$He beam will lead to several orders of magnitude increase in statistics, providing tens of thousands of $4n$ events, while the absence of low-lying $^{4,5,6}$H resonances should allow for the observation of a direct $4n$ decay, and the eventual correlations within.

\begin{figure}[t] \begin{center}
  \psfig{file=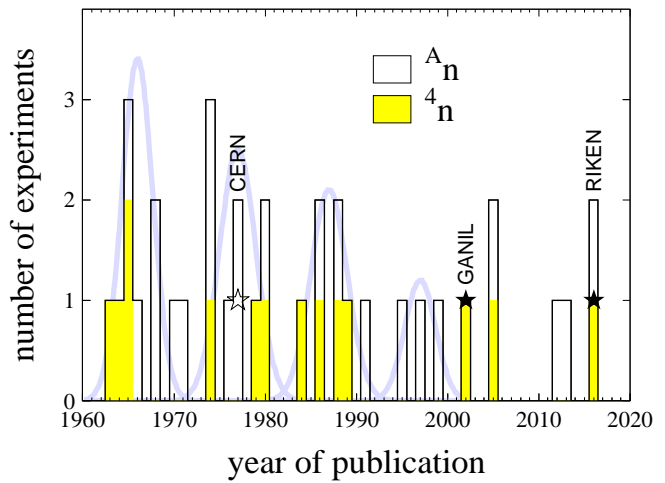,width=8.5cm}
 \end{center} \caption{The solid histogram represents the number of experiments reviewed that were searching for multineutrons (\edit{36 total}, mainly for tri- and tetraneutrons) as a function of the year of publication of the results. In yellow are those searching specifically for the tetraneutron.
 The stars represent the three positive signals reported, the empty one that was refuted \cite{Detraz77_68n_pW} and the two solid ones that have not been contested yet \cite{Marques02_4n_recoil,Kisamori16_4n_DCX}.
 The pale Gaussians guide the eye through the recurring pattern.} \label{f:history}       
\end{figure}

\subsection{A sixty-year quest}

 In this section we have reviewed experiments taking place over a sixty-year period, all searching for evidence of multineutron existence. Their chronology (Fig.~\ref{f:history}) exhibits several trends.
 Even if the techniques have been diverse and their sensitivity has increased with time, we can see a recurring pattern of `bunches' during the first forty years, with experiments accumulating from mid to end of each decade. Some experiments at mid-decade triggered others, and then the overall negative results lead to a stop in the program, until someone else restarted it a few years later.
 Towards the end of the century the number of experiments in each bunch decreased, showing signs of exhaustion due to the lack of positive signals.
 
 The number of experiments in the present century has been much lower, although two positive signals of a bound or low-lying resonant tetraneutron were obtained. 
 As said in the introduction those signals have renewed the interest in the field, both experimentally and, and as we will see in the next section, theoretically. Therefore, for the next extension of Fig.~\ref{f:history} we expect an upcoming significant `bunch' of results. Taking into account the increasing accuracy and sensitivity of these new experiments, the next few years will possibly see the end of this quest whatever the outcome, at least with respect to $^{3,4}$n.

 Moreover, besides the aforementioned experiments aiming at the detection of four neutrons already carried out \cite{NP_Kondo,NP_Marques}, complementary missing-mass experiments are also being programmed at RIKEN, without neutron detection but with increased sensitivity.
 The tetraneutron has already been revisited using the same reaction as in 2016, $^4$He$(^8$He$,^8$Be$)^4$n, with several improvements in the experimental conditions \cite{NP_Shimoura}. It has also been probed in the knockout of an $\alpha$ particle off $^8$He at backward angles in quasi-free conditions, H$(^8$He$,p\alpha)^4$n \cite{NP_Paschalis}, avoiding the FSI of the $4n$ with the other particles in the reaction.
 \edit{There are also new plans to probe the trineutron, with the reaction $^3$H$(t,^3$He$)^3$n \cite{NP_Miki}.}
 
 Among those missing-mass experiments, some are already searching for the next heavier system, the hexaneutron. One has already been carried out, knocking out two $\alpha$ particles from $^{14}$Be in the reactions H$(^{12,14}$Be$,p\alpha\alpha)^{4,6}$n \cite{NP_Beaumel}, and a second one is planned in a near future, knocking out an $\alpha$ particle and a proton from $^{11}$Li in the reaction H$(^{11}$Li$,pp\alpha)^6$n \cite{NP_Nakamura}. Depending on their results, future experiments could be planned in order to study neutron correlations in the decay of hexaneutron states.

 However, the hexaneutron seems to represent a mass frontier difficult to cross in the laboratory. In the next section we will see how theoretical calculations have been dealing with the lightest multineutrons, and how they could help us go beyond $A=6$ in order to understand possible binding energy trends in these systems.

\section{Theoretical calculations}\label{Sec_Th}

 The theoretical interest in bound or resonant multineutron states is as old as nuclear theory, and its development goes in parallel with the technical possibility to obtain accurate quantum mechanical solutions of few interacting particles using reliable interactions.
 This is an essential ingredient when dealing with systems very close to, or above, their dissociation threshold.
 The progress in this domain has been slow. The first rigorous formulation of the three-body problem in Quantum Mechanics dates from 1960 \cite{Fad_JETP39_1960}, and the first realistic solution took still several years \cite{LC_SYMPHONY_FBS_2019}.
 As it has been the case in the experiments, the main efforts of the theory have been concentrated on $^{3,4}$n, and to a lesser extent on the next heavier systems, $^{6,8}$n.

 Even if we can trace the origin of both theoretical and experimental programs back to the 1960s, it seems clear that the theoretical community has significantly increased its interest in $3n$ and $4n$ systems following each of the two experimental signals described in the previous section.
 We will thus review the many theoretical works from that perspective. We will describe the activity prior to GANIL results on the breakup of $^{14}$Be into $^{10}$Be$+4n$ \cite{Marques02_4n_recoil}, then review the different approaches that were triggered by this result, and finally discuss the many calculations that followed the second signal in the DCX reaction $^4$He$(^8$He$,^8$Be$)4n$ \cite{Kisamori16_4n_DCX}.
 But before moving to the specific works, let us start with some general considerations about the theoretical treatment of the few-neutron problem.

\subsection{The main difficulties}

 The theoretical study of $^A$n systems has faced two hard-wearing difficulties.
 The first one is finding a reliable method to solve exactly the $A$-body problem ($A=3,4...$) for loosely-bound states as well as for resonant states, which could be embedded in the continuum and very far from the physical region.
 By ``exactly'' we mean a rigorous mathematical formulation of the quantum mechanical problem and a uniformly convergent numerical scheme which could lead to the required accuracy.
 This implies an {\it ab initio} solution, i.e.\ only in terms of the $nn$ interaction.
 We anticipate that for such diluted systems the three-nucleon forces (3NF) should play a negligible role.

 Since such systems, if they manifest at all in nature, would be a subtle balance between the attractive $nn$ interaction and the Pauli `repulsion' 
and live in the continuum, any uncontrolled approximation, in the antisymmetry or in the boundary conditions, could be fatal and lead to illusionary conclusions.
 This constraint delayed the first serious attempt to study the $3n$ system to the end of the 60s, by Mitra \etal\ \cite{Mitra_PRL16_1966}, and this was still based on an ad-hoc method they had developed for solving the Schrodinger equation with a rank-1 separable interaction, which furthermore turned to be not very realistic.

 The second difficulty is related to the fact that, until the contrary is proven, $^A$n has no bound subsystems. Therefore, the usual scattering theory with two-cluster asymptotic channels cannot be applied to obtain the scattering amplitude and eventually the location of the corresponding poles.
 The only pertinent quantity, the amplitude for the $3\to3$ or $4\to4$ scattering processes at positive energy (with respect to the corresponding thresholds), is well defined but very difficult to compute, even for the simplest cases. Moreover, it is not clear that it could provide some information about possible resonant states when they are far enough from the real axis.

 Last but not least, we would like to point out that the very concept of ``resonance'' is far from being clear, and in any event it is certainly not shared by the physics community (experimentalist and theorists from many fields), even at the two-body level.
 As we will show, it is possible to generate extremely sharp `resonant' cross-sections without any dynamical near-threshold state counterpart.
 Therefore, associating a `state' to a sharp structure in the cross-section could be a dangerous shortcut.
 Moreover, the quite widely adopted definition of resonance as a rapid increase of a phase-shift crossing $\pi/2$ is not very satisfactory.
 First, it is not clear what ``rapid'' means and, most importantly, a slight modification in the interaction could significantly lower the value of the phase-shift and make a resonance `disappear', a bit surprising when dealing with a physical state.
 Not to talk about all the complexities appearing in a coupled-channel problem.

 With all that in mind, it seems that the only unambiguous approach to conclude about the existence of a $^A$n resonance is determining the location of the relevant $S$-matrix pole, or any equivalent quantity like the $A$-neutron Green function, in the complex-energy manifold (or complex-momentum plane).
 This allows to define the intrinsic parameters of the resonance in the usual way:
\begin{equation}\label{E}
 E = E_R - i\,\frac{\Gamma}{2}
\end{equation}
 whatever its proximity to the physical axis and independently of any phase-shift it can create anywhere.

\begin{figure}[t] \begin{center}
  \psfig{file=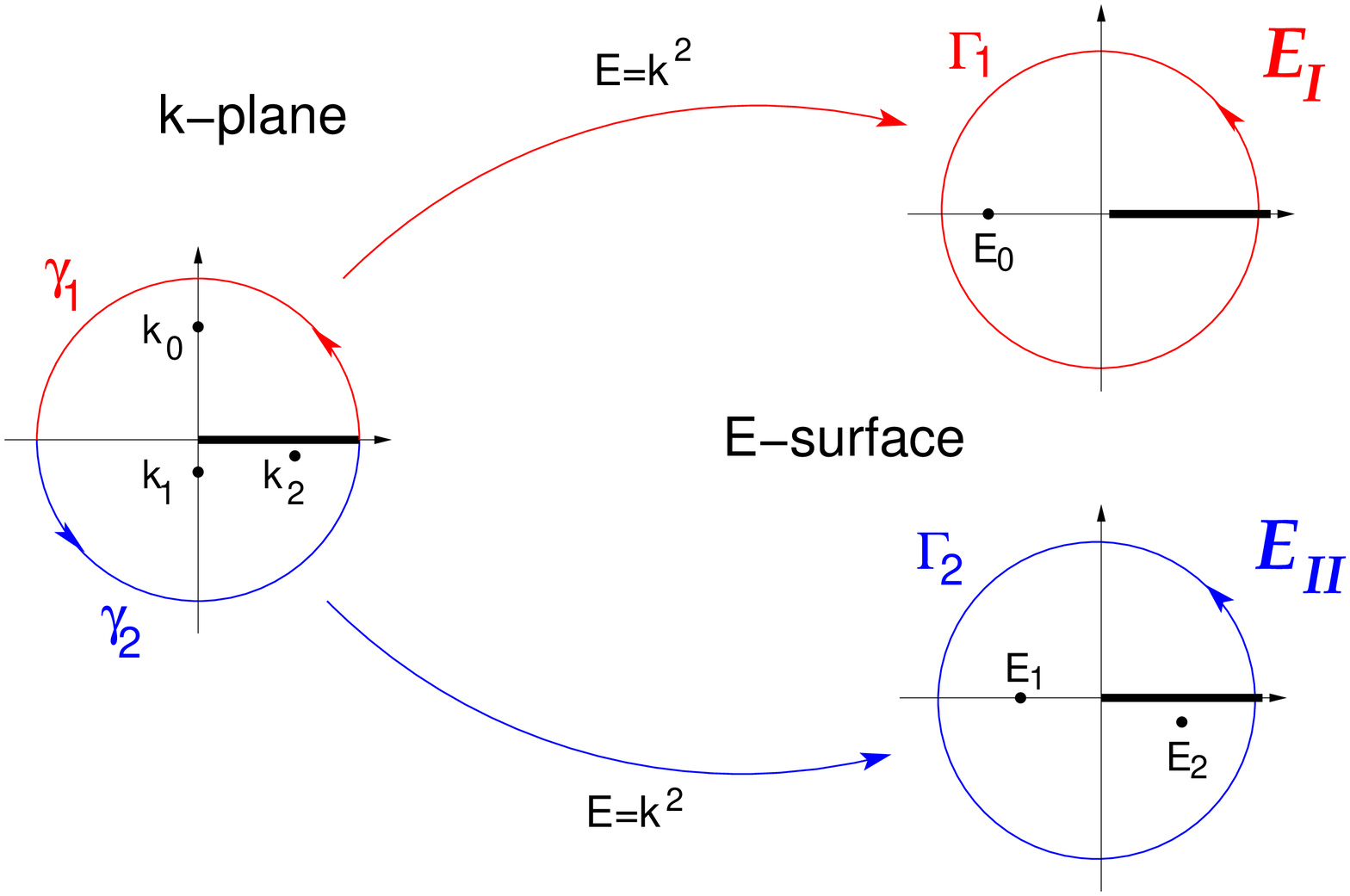,width=8.5cm}
 \end{center} \caption{Complex momentum plane and energy manifold.}\label{K_E_Plane}       
\end{figure}

 To this aim, it could be worth reminding the location of the singularities in the complex-momentum plane and in the corresponding energy Riemann surface. This is sketched in Fig.~\ref{K_E_Plane} for the simplest case of a two-cluster problem.
 In the left panel a bound ($k_0=+i\mid\!k_0\!\mid$) and a virtual ($k_1=-i\mid\!k_1\!\mid$) state appear as singularities in the imaginary $k$-axis, in the upper (bound) and lower (virtual) half planes.
 They are mapped onto the two Riemann sheets $E_I$ and $E_{II}$ of the $E$-surface, in the right panel, respectively at the points $E_0\in E_I$ and $E_1\in E_{II}$.
 Note that the energies of the virtual states are negative, as are for bound states, despite being in the continuum. 
 The upper half $k$-plane is mapped onto the first (``physical'') Riemann sheet $E_I$, and the lower one onto the second (``unphysical'') Riemann sheet $E_{II}$, which are glued to each other by the positive real axis $E>0$.

 Resonances manifest as poles in the lower half $k$-plane, $k_2=k_R-i k_I$ with $k_I>0$, and are mapped onto $E_2$ in the unphysical sheet $E_{II}$.
 They appear always by pairs ($k_2,k'_2=-k^*_2$) symmetric with respect to the imaginary $k$-axis.
 Note that Re$(E_2)>0$ only if Im$(k_2)<\,$Re$(k_2)$.
 The situation is more complicated in presence of open channels with open thresholds as well as dynamical cuts \edit{\cite{Resonance_CC_1982}}, but what is described above is enough for our purposes.
 The interested reader can find a more detailed description in the standard books of scattering theory \cite{BZP_1971,Taylor_1972,GW_1975,Newton_1982,Kukulin_1989}.

 Whether or not this $^A$n singularity could manifest in any observable, and in which way, must be considered as a different (and difficult) problem,
especially taking into account that such states can be produced in complex nuclear reactions that could mask, generally in an unknown way, the intrinsic parameters ($E_R,\Gamma$) of an eventual resonance.

\subsection{Multineutron states in the continuum}

 The research for an eventual bound state is relatively easy. If a direct bound state calculation gives a negative eigenstate of the Hamiltonian, the system is bound. On the contrary, we conclude that there is no bound state... within the accuracy of the method, which provides then a lower limit of the binding energy.
 The method for computing bound states must be accurate enough to deal with loosely bound systems, i.e.\ quite extended objects in configuration space.
 We would like to point out from now that no serious calculation has ever predicted a $3n$ or $4n$ bound state.

 However, it is worth noting that the aforementioned pioneering calculation of Mitra \etal\ \cite{Mitra_PRL16_1966} was quite optimistic. 
 They realized that a pure S-wave $nn$ interaction generates a strong repulsion in the $3n$ system due to the Pauli principle, but concluded that ``a moderately attractive $^3$P force ... is enough to produce a bound $^3$n state'', and added that the ``range and strength of the force seem to be in good accord with the $^3$P$_0$ phase-shift data of Bryan and Scott'', one of the most realistic potentials available at that time. Their conclusion, certainly due to the simplicity of their dynamical approach and interactions, would be denied very soon.
 
 In the absence of bound states, the optimal choice is to find the location of the nearest singularity in the continuum, i.e.\ the parameters ($E_R,\Gamma$) of a virtual or resonant state with all the caveats mentioned.
 This is however a difficult task when working in configuration space, mainly due to the exponentially increasing boundary conditions.
 The `regalian' way to avoid the latter is the Complex Scaling Method (CSM) \cite{Nuttal_PR188_1969,CSM-ref1,CSM-ref2}. In the two-body case it consists of an analytic continuation of the Schr\"odinger equation by means of a complex rotation of the coordinates
\edit{($r\to r\, e^{i\theta}$)},
 which \edit{transforms} the exponentially increasing boundary conditions of a resonance into square integrable ones, provided the ``rotation angle'' $\theta$ is larger than half the argument \edit{$\theta_r$} of the complex resonance energy (\ref{E}):
\[ 0< \theta - {\theta_r  \over 2}  < {\pi\over 2}  \qquad \tan\theta_r= {\Gamma\over 2E_R}  \]

\begin{figure}[t] 
\begin{center}
  \psfig{file=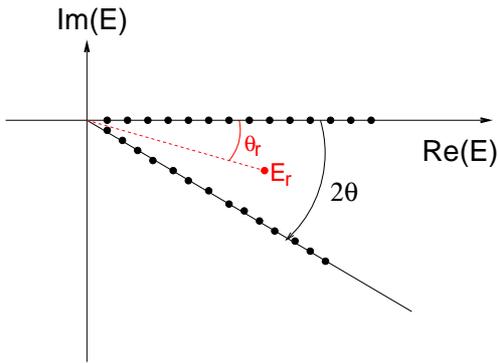,width=6.5cm}
 \end{center} \caption{The principle of the Complex Scaling Method.}\label{FIG_CSM}       
\end{figure}

 Under these conditions, illustrated in Fig.~\ref{FIG_CSM}, the resonant state can be computed by means of the standard bound-state problem techniques.
 This approach, supplemented with many variants, has been successfully extended to $A=3,4,5$ \cite{Nuttal_PR188_1969,Aoyama_PTP116_2006,LC_PRC84_2011,Rimas_PRC86_2012,LC_FBS54_2013,Myo_2014}.
 Note, however, that the numerical difficulty increases dramatically with the rotation angle, in turn limited by the analytical cuts of the two-body interaction, and this restricts the practical application of CSM to relatively narrow states \cite{Rimas_PRC86_2012}.

\begin{figure}[t] 
\begin{center}
  \psfig{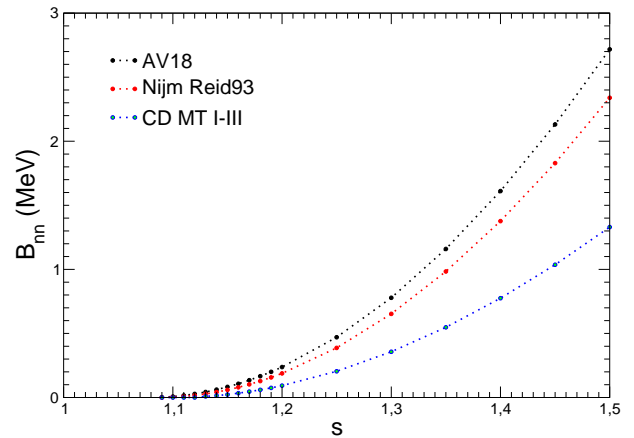}
 \end{center} \caption{Scaling factors (\ref{V_s}) needed to bind $2n$ in the $^1$S$_0$ state for some $nn$ interactions. AV18 and Reid93 give $s\gtrsim1.08$.} \label{FIG_B_s}       
\end{figure}

 To circumvent this problem, one possibility is to introduce an additional binding to the physical system under study until it gets artificially bound. One can then remove adiabatically the `artifact' and follow the evolution of the system in the continuum until it reaches the physical state.
 A commonly used artifact consists of a ``scaling factor'' $s$ in front of the $nn$ interaction, which can eventually depend on each partial wave $\alpha\equiv\{L,S,J\}$:
\begin{equation}\label{V_s}
 V^\alpha_{s_\alpha}(r) = s_\alpha\, V^\alpha_{nn}(r)
\end{equation}
and increase $s$ until the multineutron system binds.
 Note that only moderate $s$ values are required to bind the dineutron with all realistic and semi-realistic $V_{nn}$ models, as shown in Fig.~\ref{FIG_B_s}.
 For instance, the critical value to get $B_{nn}>0$ is $s=1.080$ using the Argonne AV18 potential \cite{AV18_1995} and $s=1.087$ using Nijmegen Reid93 \cite{Reid93}.
 For the phenomenological CD-MT13 \cite{MT_NPA127_1969,HLMC_PRCRC_2019}, 
 it is only slightly larger, $s=1.101$, probably due to its shorter range.

 Another possibility consists in placing the multineutron system inside an external potential, which acts as a confining trap. This is implemented by adding to the total potential energy $V$ of the system a term of the form:
\[ V=\sum_{i<j=1}^A V_{ij} \ \ \rightarrow\ \ V=\sum_{i<j=1}^A V_{ij} + \sum_{i=1}^A V_T(r_i) \]
where $V_{T}$ can be an harmonic oscillator (HO) or a Woods-Saxon potential.
 The quantum mechanical solution of $2n$ and $3n$ systems in a Woods-Saxon trap is treated with some detail in the Appendix.
 
 The greatest advantage of this approach is the preparation of the system in an {\it a priori} well-controlled bound state. Moreover, the size of the extra energy required to bind the system constitutes already a first and valuable indication of how far from the physical region a resonant state may be.
 The main drawback, however, is that the artifact can dramatically change the properties of the system 
 and induce some parasitic phenomena that have no physical reality. 
 This unpleasant situation may happen if one makes use of the two  following procedures.

 The first one is the introduction of spurious processes that can modify the evolution of the system.
 For example, if one artificially binds $^3$n by introducing a scaling factor $s$ in the $nn$ interaction, the dineutron will be quickly bound
by several tens of MeV. For $s\approx3$, a typical value required to bind $^3$n, $B_{nn}\approx100$ MeV.
 Therefore, a multineutron state built in such a way will contain one or several subsystems with very peculiar properties (e.g.\ extremely compact), that will certainly affect its evolution into the continuum.
 What one believes to be a bound $^3$n state will be in fact a system strongly decaying into $n+^2$n.

 If the method used to solve the three-body problem does not allow to properly account for open scattering channels, like when using basis of square integrable functions, this fact will not be detected in the numerical calculations but will influence the properties and the evolution of a state bound by brute force in an uncontrolled way.
 In any event, the evolution of the multineutron system when decreasing $s$ will be totally dominated by this process and one may attribute to $^3$n  resonances what corresponds to an unphysical decay $^3$n$\,\to n+^2$n.
 The same will happen when studying the tetraneutron with the corresponding spurious channels $^4$n$\,\to n+^3$n and $^4$n$\,\to ^2$n$+^2$n.

 The appearance of these spurious thresholds can be avoided by adding an $A$-neutron force involving only the total number of neutrons of the studied system. For example, for $^3$n we can introduce a $3n$ force that we can choose for simplicity of hyper-radial form: 
\begin{equation}\label{W}
 W_{3n}(\rho) = -W_0 \, f (\rho)    \qquad  \rho^2=x^2+y^2 
 \end{equation}
 where $f(\rho)$ is an asymptotically decreasing function of $\rho$ (like $e^{-\rho}$ or Yukawa) and ($x,y$) are the intrinsic Jacobi coordinates. This artifact does not modify any subsystem threshold and the evolution $W_0\to 0$ can be done safely.

\begin{figure}[t] 
\begin{center}
  \psfig{file=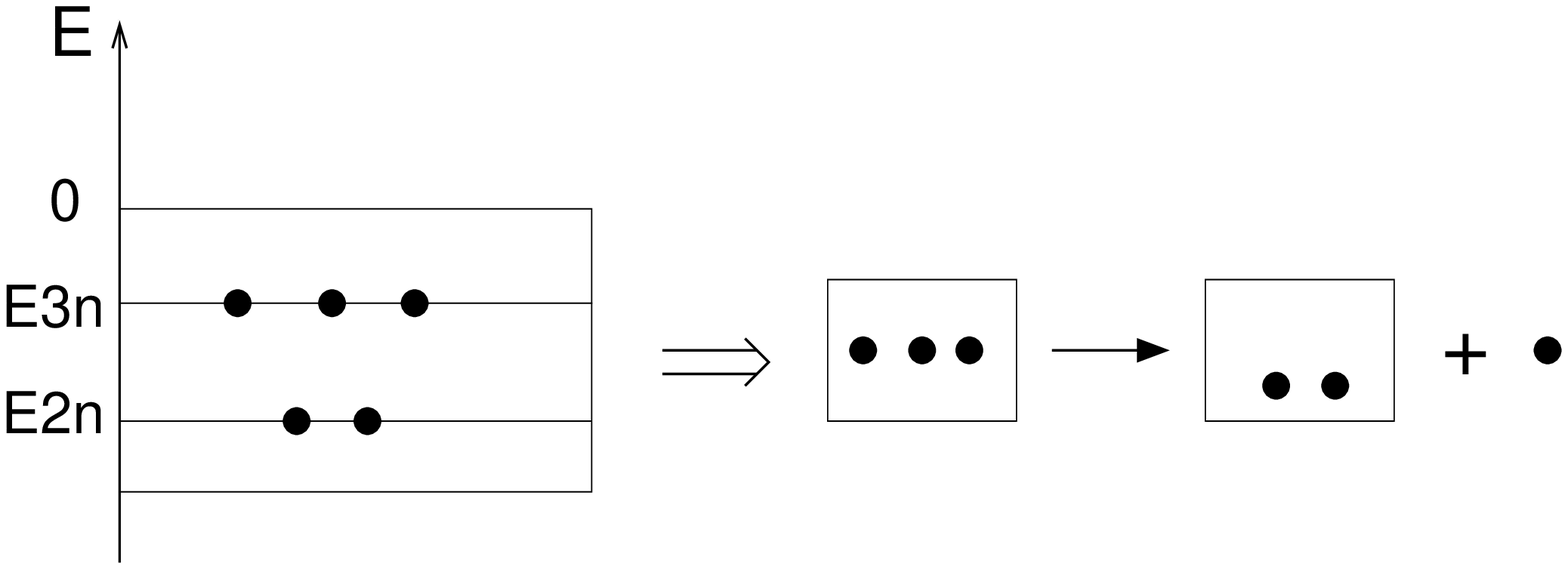,width=7.5cm}
\vspace{0.5cm}  
  \psfig{file=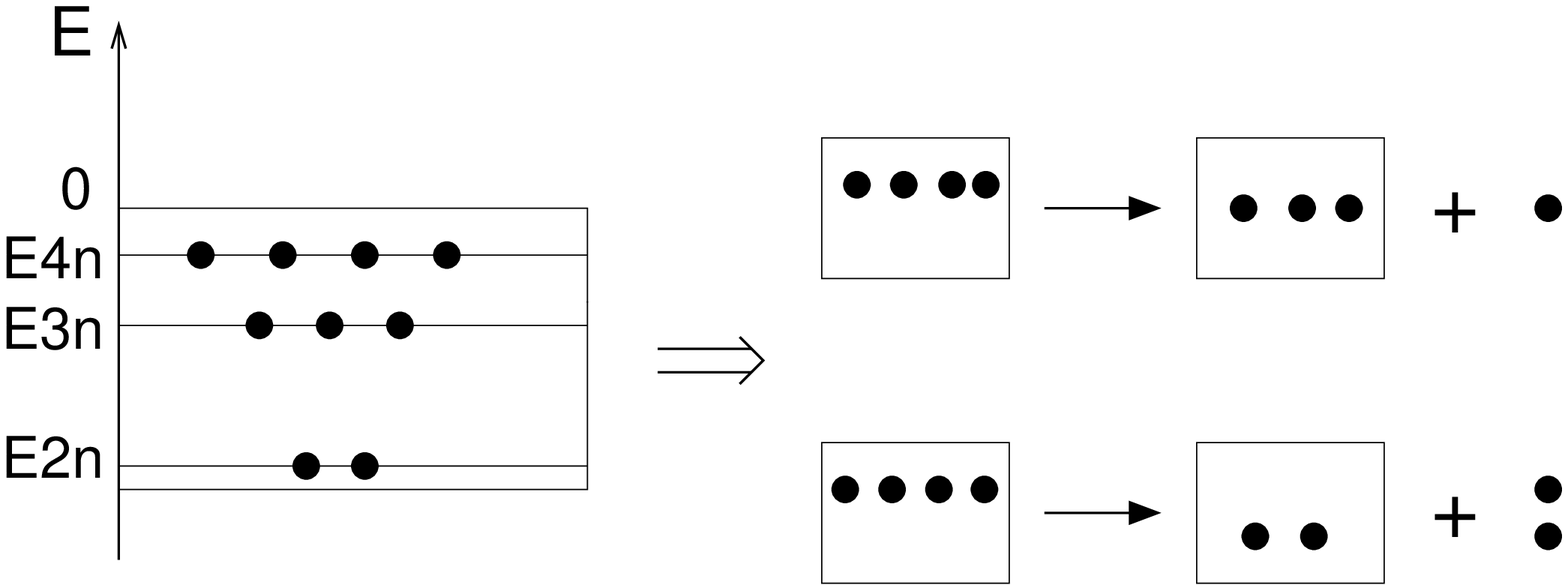,width=7.5cm}
 \end{center} \caption{Possible decay channels of $3n$ and $4n$ in a trap.} \label{Trap_3n}       
\end{figure}

 A similar distortion is also induced when `confining' neutrons in a Woods-Saxon trap, like for example:
\begin{equation}\label{V_T}
 V_T(r) = \frac{V_0}{1+e^{(r-R)/a}}
\end{equation}
 a common practice in the research on neutron droplets \cite{Ndrops1,Ndrops2,Ndrops3}.
 For small multineutrons, the evolution of their binding energies as a function of the potential strength $V_0$ has been used to draw conclusions about eventual $3n$ and $4n$ resonances.
 However, this approach presents some instabilities, as we illustrate for the simplest case of $3n$ in the upper panel of Fig.~\ref{Trap_3n}.
 If $B_{2n}>B_{3n}$ for a given set ($V_0,R,a$) of the confining potential (\ref{V_T}), the $^3$n will decay into a dimer, staying in the well, plus 
an outgoing neutron with kinetic energy proportional to $B_{2n}-B_{3n}$. 
 If the methods used to compute the bound state account for the full quantum dynamics of the system, the supposed confined multineutron state must decay. 
 This fact may strongly impact the extrapolation pattern. Since  the presence of thresholds is associated with  algebraic branching points, 
 the open decay channels will generate sharp structures in the extrapolation function and the results of any extrapolation procedure become unreliable.
 The binding energy of multineutrons in a trap is in fact only fictitious, as are the properties extrapolated from their evolution as a function of the trap parameters.
 The same will happen for a $^4$n state, with respect to the open thresholds $n+^3$n and $2n+^2$n (Fig.~\ref{Trap_3n}, lower panel).

 The second mechanism that may induce parasitic phenomena is related to the evolution 
 from a bound state to a state in the continuum.
 It is not a mere extrapolation from negative to positive values in an energy axis. As already noted, the quantum mechanical energy continuum is not an axis, even not a plane, but a Riemann manifold with dynamical
cuts and delicate analytic properties.

 The evolution of a bound state into the continuum can be properly described by the so-called Analytic Continuation on the Coupling Constant Method (ACCCM) \cite{Kukulin_1989}.
 It consists in adding to the Hamiltonian $H$, for which we aim to determine the parameters of a resonance, a perturbation $\lambda W$.
 We can then make use of the analytic dependence of the spectrum of the `perturbed' Hamiltonian
$H_{\lambda}= H + \lambda W $ on the coupling constant $\lambda$.
 If $k$ is the wave number of a bound state with negative energy $E$ of $H_{\lambda}$ ($E=\hbar^2k^2/2\mu$), its dependence on $\lambda$
is given by the Pad\'e-like analytic expansion:
\begin{equation}\label{k_NM} 
k(\lambda)=  i {P_N(z)\over Q_M(z)} =  i \frac{ c_0+c_1 z + c_2 z^2+ \ldots c_Nz^N}   { 1+d_1 z + d_2 z^2+ \ldots d_Mz^M }   
\end{equation}
where $P_N$ and $Q_M$ are polynomials on the complex variable $z=\sqrt{\lambda-\lambda_c}$.
 The critical value $\lambda_c$ corresponds to a zero-energy bound state that moves into the continuum
\edit{and to an algebraic branching point of the complex function (\ref{k_NM})}.

\begin{figure}[t] 
\begin{center}
  \psfig{file=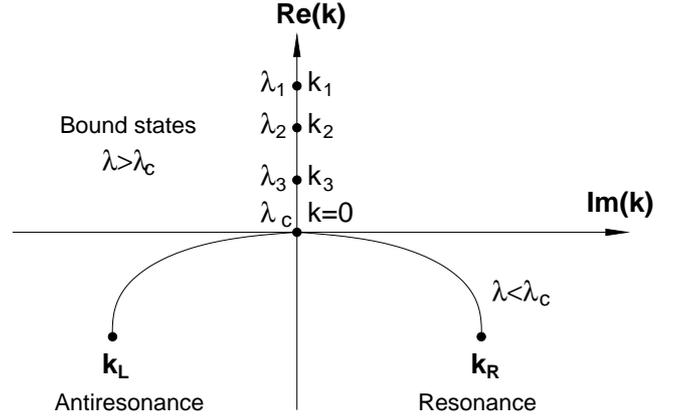,width=8.5cm}
 \end{center} \caption{\edit{Schematic view of the ACCC method. An ensemble  of bound state positions ($\lambda_n,k_n$) is computed for $\lambda>\lambda_c$, a critical
branching point where $k=0$, and the trajectory of the $S$-matrix resonance pole for $\lambda<\lambda_c$ is determined.}} \label{FIG_ACCC}       
\end{figure}

 ACCCM \edit{is schematically illustrated in Fig.~\ref{FIG_ACCC} and} proceeds as follows. 
 In a first step, a series of bound state values ($\lambda_n,k_n$) are computed, with $\lambda>\lambda_c$ and $k_n$ purely imaginary. They are inserted in the formal expansion (\ref{k_NM}) to determine the $N+M$ unknown coefficients ($c_i,d_i$), which are real.
 The $k(\lambda)$ dependence being fixed, one can decrease \edit{$\lambda\to 0$} to determine the resonance position of $H$ in the complex plane. Note that for $\lambda<\lambda_c$, $z$ is a complex number, as will be $k$ and the corresponding $E$.
 This method has been successfully used to compute the position of the nearest $3n$ and $4n$ resonances \cite{LC_3n_PRC71_2005,LC_4n_PRC72_2005,HLCK_4n_PRC93_2016}, as well as of the $^{4,5}$H isotopes \cite{LHC_PLB791_2019}. It requires an accurate determination of the bound state ($\lambda_n,k_n$) parameters and a careful stability study as a function of the degrees $N$ and $M$ of the Pad\'e expansion (\ref{k_NM}).

 After this overview of the general issues related to the calculation of few-neutron systems, in the following we will review the most relevant theoretical works that have been published prior, in between and after the two experimental signals observed at GANIL in 2002 and RIKEN in 2016.

\subsection{Previous to GANIL 2002}

 Before 2002 we find in the literature several works devoted to $3n$ and $4n$ systems which, although far from being uninteresting, can be considered as preliminary or incomplete from the physical point of view, either due to the quantum mechanical treatment of the three-body problem and/or to a very simplistic $nn$ interaction.
 As we noted in the introduction, the first result to our knowledge is from Mitra \etal\ back to 1966 \cite{Mitra_PRL16_1966}, using a technique for solving the three-body Schr\"odinger equation with rank-1 separable S- and P-wave $nn$ interactions. 
 Their optimistic conclusion about a bound $^3$n was immediately contested by Okamoto \etal\ \cite{Okamoto_Davies_PLB24_1967} and Barbi \cite{Barbi_NPA99_1967}, showing with variational approaches that based on their knowledge of the NN interaction the trineutron was unbound by at least 10~MeV, and that the depth of the required $nn$ potential well to bound it should be unphysically large.

 The first `exact' $3n$ calculation solving the Faddeev equations in momentum space, although using a rank-1 separable Yamaguchi force, is due to Gl\"ockle in 1978 \cite{Glockle_3n_PRC18_1978}. 
 It is also the first time that the trajectory of the complex-energy $^3$n resonance as a function of a scaling factor $s$ was presented, together with a careful discussion of the analytic continuation properties of the scattering amplitude.
 The critical value for binding $3n$ was found to be $s=4.2$, and the possibility of a low-energy resonance was ruled out.
 Of course the latter was on the basis of a pure S-wave interaction, but for an educated practitioner of the few-nucleon problem it was clear that if such a huge extra binding was needed, the P-wave would not be enough.

 The severe dynamical limitations of the first Gl\"ockle paper were partly solved one year later together with Offermann \cite{OG_3n_NPA318_1979}, using S+P waves and the Reid potential including a tensor force.
 They found that when adding attractive P-waves in $V_{nn}$, the critical S-wave scaling factor to bind $3n$ was slightly lowered to $s=3.7$, but still remained unacceptably large.
 They also determined that when using a realistic $nn$ interaction the most favorable candidate was $J^{\pi}=3/2^-$, dominated by a $^3$PF$_2$ force
but with a scaling factor of $s\approx 4$. The other candidates were $J^{\pi}=1/2^-,3/2^+,1/2^-$, by order and in contradiction with Mitra \cite{Mitra_PRL16_1966}, who had found the positive-parity state the most favorable one.
 Despite all these refinements they reached, however, the very same conclusion: no chance for a $3n$ bound state and a ``low energy resonance is extremely unlikely''.
 They ended with a non-prophetical remark:
 ``The [required modifications in $V_{nn}$] are not so small that one likes to push forward a very expensive experiment to measure a three-neutron resonance''. Would it be enough to discourage any further experimental or theoretical research? Certainly not!

 We are aware of other works that appeared during the following decade \cite{Sunami_PGTh63_1979,Bevelacqua_NPA431_1980,BBKE_SJNP41_85,Gorbatov_YF50_89,Moller_Orlov_SJPN20_1989}, but we are not going to comment on them with further detail for they represent still a preliminary stage, due to the interactions and/or the techniques used to solve the three-body problem.

 At the end of this period, two works devoted to $3n$ and $4n$ resonances were published, which deserve special attention for they were also based on exact solutions of the three- and four-body problem.
 
 In 1997 Sofianos \etal\ \cite{SRV_JPG23_1997}, within the framework of the hyper-spherical formalism and CSM, computed the position of the $3n$ and $4n$ resonances in the complex-momentum plane, identified with the zeros of the Jost function or 
 the $S$-matrix poles.
 The $nn$ interaction was limited to two different central S-wave potentials, one purely attractive Yukawa and the semi-realistic  MT13 \cite{MT_NPA127_1969}. 
 With the MT13 potential, 
 the lowest-lying $3n$ resonance was $E=-4.9-6.9\,i$~MeV with $J^{\pi}=1/2^-$ (degenerated with $3/2^-$ within the considered potentials), and it was located in the third quadrant of the second Riemann sheet (Fig.~\ref{K_E_Plane})\footnote
{Note that there is a misprint in Table 1 of Ref.~\cite{SRV_JPG23_1997},  the real part of the resonance energy appears with opposite sign.}.
 The search for a $4n$ resonance with the hyper-spherical method gave for the nearest state $J^{\pi}=0^+$ and $E=-2.64-8.95\,i$~MeV, i.e.\ also located in the third quadrant of the second Riemann sheet.

 Although limited by using the ``minimal approximation'' in the hyper-spherical expansion and a semi-realistic NN potential, the conclusion from this work was clear. First, any $3n$ and $4n$ possible bound state is very far, since one requires an unphysically large enhancement factor in the NN interaction (typically $s\approx 4$ and 3.6 respectively for $3n$ and $4n$ with the MT13 potential), well beyond any possible uncertainties of the models used.
 Second, the resonant states, although well identified, are located in the unphysical region with no possibility to manifest in an scattering experiment.

\begin{figure}[t]
\begin{center}
  \psfig{file=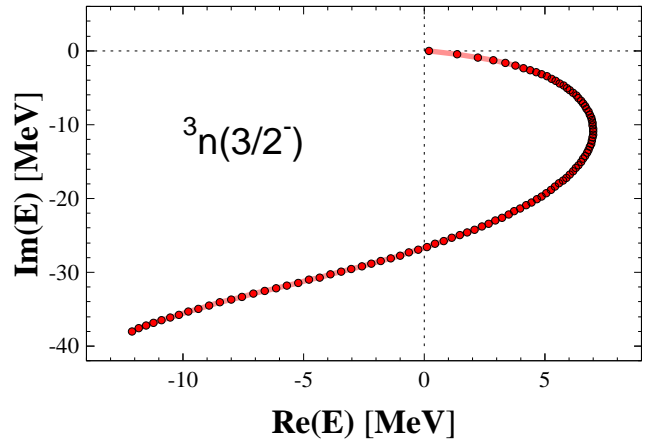,width=8.5cm}
\caption{$S$-matrix pole trajectory of $3n$ in $J=3/2^-$ with a rank-2 separable potential. Adapted from Ref.~\cite{HGK_3n_PRC66_2002}.} \label{3n_Gloeckle_32m}       
\end{center}
\end{figure}

 In 1999 Witala and Gl\"ockle \cite{WG_3n_PRC60_1999,Witala_FBS13_2001} published a solution of the Faddeev equations in configuration space for the $3n$ system.
 The $nn$ interaction was taken from the phenomenological GPdT model \cite{GPdT_NN_PLB32_1970} and from the realistic Reid93 Nijmegen potential \cite{Reid93}, which includes the $^1$S$_0$, $^3$P$_0$ and $^3$PF$_2$ channels, all attractive.
 They introduced a different scaling factor $s_{\alpha}$ in each partial wave of the $nn$ potential, except in $^1$S$_0$ to keep the dineutron unbound. They first forced the $3n$ system into a bound state and decreased $s_{\alpha}$ progressively following the resonant trajectory in the complex-momentum plane.
 The $J^{\pi}=1/2^-,3/2^-,3/2^+$ 
 resonance positions were obtained by using the so-called Smooth Exterior Complex Scaling Method \cite{SECSM}.

 Due to numerical instabilities they could not reach the physical case with all $s=1$. Their conclusion was however similar to the previous works: the $3n$ resonance energies will have such a large imaginary part that one should not expect to see any trace of them in experiments.

\subsection{From GANIL to RIKEN}

 In 2002, few months after GANIL publication of the $^4$n signal \cite{Marques02_4n_recoil}, there was another paper from Gl\"ockle \etal\ concerning the $3n$ system \cite{HGK_3n_PRC66_2002}.
 They used rank-2 separable $nn$ forces ($^1$S$_0$, $^3$P$_0$, $^3$P$_2$, $^1$D$_2$) and solved the Faddeev equations, which were analytically continued into the unphysical energy sheet below the positive real-energy axis.
 The trajectories of the $3n$ $S$-matrix poles in the $J=1/2^\pm$ and $3/2^{\pm}$ states were traced as a function of the scaling factors of the $nn$ forces (Fig.~\ref{3n_Gloeckle_32m}), but only in the $^3$P$_0$, $^3$P$_2$ and $^1$D$_2$ channels in order to keep $2n$ unbound.
 The final positions of the $S$-matrix poles, once the artificial factors removed, were found to be in all cases far from the positive real-energy axis, which provided a strong indication for the nonexistence of nearby $3n$ resonances.

 This was the last of a series of papers which were pioneers on this topic \cite{Glockle_3n_PRC18_1978,OG_3n_NPA318_1979,WG_3n_PRC60_1999}.
 The overall conclusion was that $3n$ resonances could not exist near the physical region.
 The clarity of these works and the convincing arguments put forward should have been sufficient to discourage once and for all any claim for a $3n$ resonance.
 As we will see, this is unfortunately far from being the case.

 Immediately after GANIL publication of the eventual observation of a tetraneutron, three consecutive papers were published \cite{Timofeyuk:2003ya,Bertulani_Zellevinsky_JPG29_2003,Pieper:2003dc} as a reaction to this surprising result.
 They agreed in the impossibility of a bound tetraneutron, based on the fact that the modifications of the NN (and eventually NNN) interaction required to obtain a $^4$n bound state were so large that the description of all the surrounding nuclear chart ($^4$He, $^4$H, $^5$He...) would be destroyed, and not by a small quantity but by several tens of MeV.
 If the conclusion was unanimous, the methods used were however quite different.

 In 2003 Timofeyuk \cite{Timofeyuk:2003ya} used \edit{Hyperespherical Harmonics  (HH) formalism and} a quite simplistic NN interaction (Reichstein and Tang \cite{RT_NPA139_1969}) to compute the lowest  diagonalized hyper-radial potential $V_{KK'}(\rho)$ in the model space $K_{\rm max}=16$, and all of them where found to be monotonously repulsive for the $2n$, $4n$, $6n$ and $8n$ systems. The possibility of a narrow resonance was also excluded in any of these systems.
\edit{In a previous preprint \cite{Timofeyuk_Preprint_2002} the same author had proposed alternative transfer reactions to investigate multineutron states and evaluated some cross section using the DWBA approximation.}
 
 Later in 2003 Bertulani \etal\ \cite{Bertulani_Zellevinsky_JPG29_2003} considered a totally different approach. They computed the interaction between two $2n$ dimers artificially bound in a simplistic NN model (Volkov \cite{VOLKOV_NP74_1965}), hoping to find a possible attraction among them, as one could expect between two bosonic-like ($^1$S$_0$) systems.
 Whatever they tried, they systematically obtained a dimer-dimer repulsion. They concluded about the impossibility to explain any $4n$ bound or resonant state within this model, ``although more complex variational approaches still can be explored'', as well as any possible $T=2$ $4n$ forces.
 Despite the simplicity of the model and the crude approximations used in obtaining the dimer-dimer potential, their conclusion turns out to be quite general as it will be discussed later in the framework of EFT \edit{(Effective Field Theory)} for fermions in the unitary limit.

 A few months earlier Pieper \cite{Pieper:2003dc} had applied Green Function Monte-Carlo (GFMC) methods to study the possible existence of a $^4$n bound state. He used the NN Argonne V18 interaction supplemented with the Illinois IL2 3NF \cite{AV18_IL2_2001}, which had been very successful in reproducing the binding energies of $A=2$--10 nuclei.
 The conclusion was the same than in previous works: any attempt to force a $^4$n bound state would have ``devastating effects'' in the description of the nuclear chart.
 However, Pieper surprisingly claimed a possible $^4$n resonance at about only 2~MeV.
 These two conclusions seemed somehow contradictory: the $4n$ Hamiltonian could hardly accommodate at the same time the absolute impossibility of a bound state and a near-threshold resonance.

\begin{figure}[t]
\begin{center}
  \psfig{file=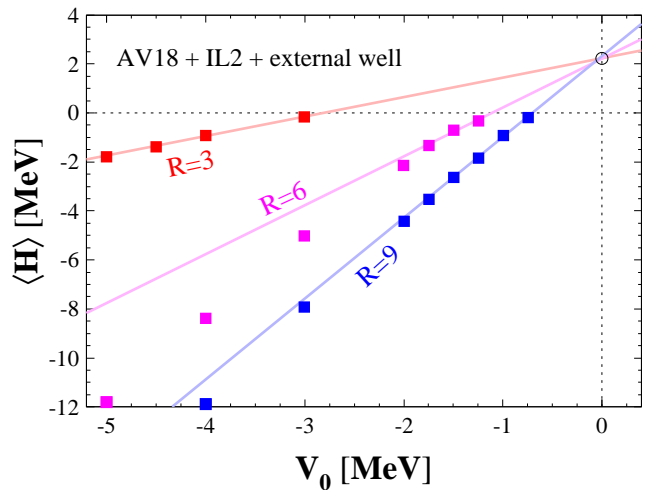,width=8.5cm}
\caption{Energies of $4n$ in a Woods-Saxon trap (\ref{V_T}) as a function of the potential strength $V_0$. The results obtained with several values of $R$ are used to extrapolate linearly into the continuum. Adapted from Ref.~\cite{Pieper:2003dc}.}\label{E4n_WS_Trap}       
\end{center}
\end{figure}

 The possible resonant state was determined by computing the energy of $^4$n bound states in a Woods-Saxon trap (\ref{V_T}) with fixed $a$ \cite{Pieper:2003dc}. Using different values of $R$, the potential strength $V_0$ was progressively decreased until the $4n$ became unbound, and the last energy values were extrapolated towards the absence of trap at $V_0=0$ (Fig.~\ref{E4n_WS_Trap}). All radii led to similar extrapolated values of $E\sim2$~MeV.
 Pieper's work had a great influence in this topic and it was taken as a definite proof to justify further experimental, and some theoretical, investigations.
 It is worth noting, however, that his own statement was very cautious: ``since the GFMC calculation with no external well shows no indication of stabilizing at that energy, the resonance, if it exists at all, must be very broad''.

 \edit{In 2004, Grigorenko \etal\ published a work devoted to $4n$ and $^5$H in the framework of the HH method \cite{GTZ_EPJA19_2004}. An interesting critical review of 
 $4n$ results can be found in Sec.~4 of that paper. They explained why no indication of $4n$ FSI was found in the pion DCX reactions, and concluded that any resonant state, if existing at all, must be very broad.
 Furthermore, they computed with a reaction model how the position of the broad $4n$ peak can be shifted depending on the reaction used to observe it. They pointed out, with illustrative examples from standard Quantum Mechanics textbooks, that sharp resonance-like structures in the continuum can be formed even if only repulsive potentials are present in the system.}

 In 2005 Lazauskas \etal, motivated by the paradoxical results of Pieper, undertook a series of works to locate the nearest $3n$ \cite{LC_3n_PRC71_2005} and $4n$ \cite{LC_4n_PRC72_2005} resonances in the complex-energy plane using a realistic $nn$ Reid93 interaction with CSM and ACCCM, adapted to properly deal with the continuum.
 The technique used for solving the three- and four-body problem was the Faddeev-Yakubovsky (FY) equations in configuration space, which were already tested in similar systems like $^4$He, $p+^3$H, $n+^3$He, $^2$H$+^2$H and $n+^3$H \cite{LC_Frontiers_2019}.
 The only approximations made were the number of grid points and the partial-wave expansion in both interactions, and the FY amplitudes which were carefully checked.
 Several $J^{\pi}$ states were examined, being the nearest ones $3/2^-,1/2^-,1/2^+$ for $3n$ and $0^+$ for $4n$.

\begin{figure}[t] 
\begin{center}
  \psfig{file=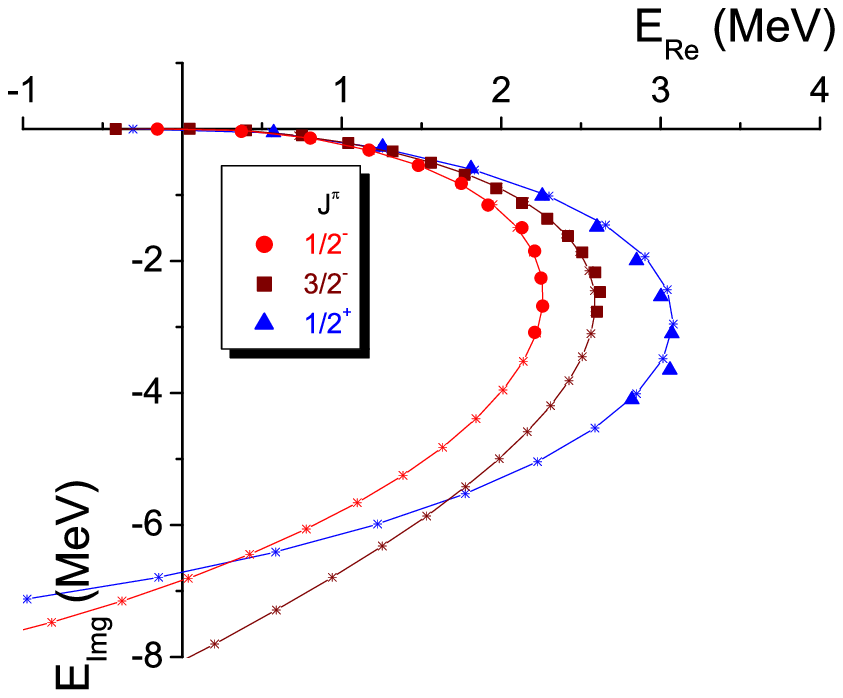,width=7.cm} \\
  \psfig{file=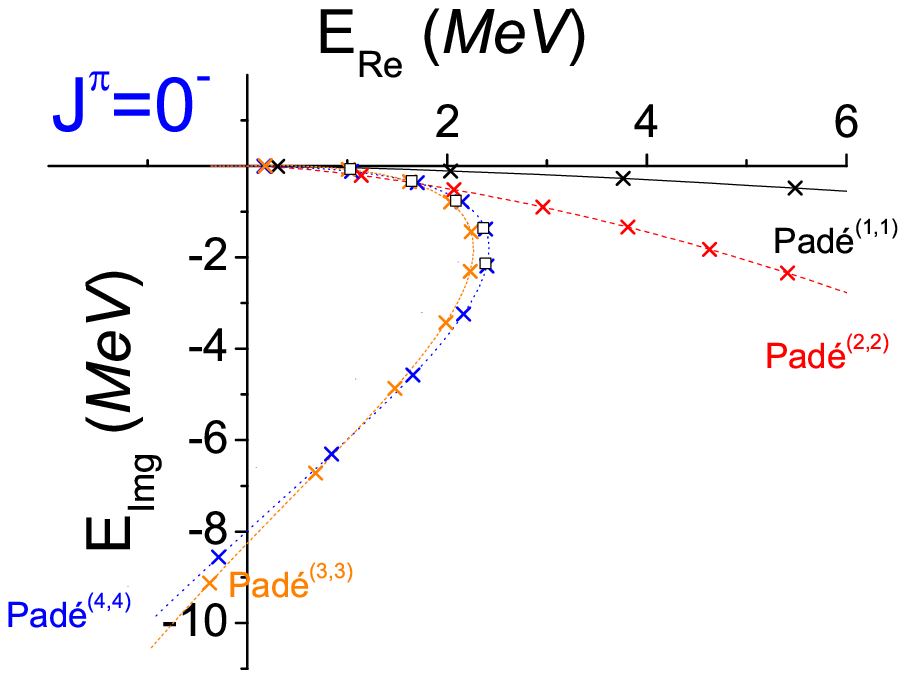,width=7.cm}
 \end{center}
\caption{Selected resonance trajectories for the $3n$ (upper) and $4n$ (lower) states as a function of the strength $W_0$ of the artifact. Results were obtained by combining ACCC and CS (solid points) methods. Adapted from Refs.~\cite{LC_3n_PRC71_2005,LC_4n_PRC72_2005}.} \label{FIG_TRAJEC_3n_4n}       
\end{figure}

The trajectories in the complex plane were plotted as a function of the strength of an artificial three- and four-body force of the type (\ref{W}),
starting from a bound state ($W_0=W_0^c$) and moving to the continuum by decreasing the strength until the physical value at $W_0=0$.
 All of them were located very far from the physical region well before $W_0$ was removed, lying in the third quadrant of the second sheet, i.e.\ Im(E)$\,<0$ and Re(E)$\,<0$, often known as ``sub-threshold resonances'' (Fig.~\ref{K_E_Plane}).
 We have displayed in the upper panel of Fig.~\ref{FIG_TRAJEC_3n_4n} the $J^{\pi}$=$3/2^-$, $1/2^-$, $1/2^+$ trajectories of $3n$ 
obtained with the CSM and ACCCM.
 The lower panel 
 shows the $0^-$ trajectory of 4n using ACCCM, where the convergence of the method as a function of the degree $(N,M)$ in (\ref{k_NM}) can be observed. 

 The conclusion was that there could be no observable $3n$ or $4n$ resonance, a fortiori bound state, with realistic nuclear forces, in close agreement with the previous $3n$ work \cite{HGK_3n_PRC66_2002} despite the different interaction and methods used. It was pointed out that Pieper's conclusions were certainly due to the extrapolation procedure used.

 For the sake of completeness, 
 we note that in 2008 Anagnostatos \cite{Anagnostatos_IJMPE17_2008} claimed that $^4$n and $^6$n could be particle-stable. 
 Within the framework of the so-called Isomorphic Shell Model, mixing semi-classical and classical mechanics, he concluded that, even if the results depended on the strength of $V_{nn}$, at least low-lying resonant states could be reasonably expected, with $^6$n being more favored than $^4$n.
 These conclusions remain however very doubtful, for the very peculiar approach used to solve the few-body quantum mechanical problem in a classical way.

\subsection{Post RIKEN 2016}

 After RIKEN publication of a new tetraneutron signal \cite{Kisamori16_4n_DCX}, the theoretical interest on this field was renewed and even enhanced. We remind that this new result was in fact also compatible with a bound state, while GANIL result was found to be also compatible with a low-lying resonance \cite{Marques05_4n_recoil}, and they were therefore in agreement.

 In the same 2016, Hiyama \etal\ \cite{HLCK_4n_PRC93_2016} published a joint effort between the authors of Refs.~\cite{LC_3n_PRC71_2005,LC_4n_PRC72_2005} and theorists from RIKEN, working independently and using different approaches for solving the few-nucleon problem.
 The work from 2005 \cite{LC_4n_PRC72_2005} let no hope for the RIKEN signal to be attributable to a $4n$ resonance, and it was natural to clarify not only the tension with the experimental results but also with Pieper's conclusion.

 The philosophy, as well as the computational methods, were different. 
 From the physical point of view,
 the strategy was to keep unchanged realistic $V_{nn}$ forces and the well controlled $V_{nnn}$ forces in the $T=1/2$ total isospin channel, and let as unique free parameter to reproduce RIKEN results the strength of $V_{nnn}$ in the $T=3/2$ channel.
 The introduction of the $T=3/2$ 3NF was required to account for the systematic underestimation by the UIX model \cite{AV18_IL2_2001} of the binding energies in the He isotopic chain,
 although it was found to be smaller than the $T=1/2$.
 The advantage of this approach was to preserve the good description by the modern nuclear Hamiltonians of all the systems not (or little) sensitive to this 3NF component.

 From the technical point of view, the $4n$ solutions in configuration space
were obtained in two different ways: by using the variational Gaussian Expansion Method \cite{Kami88,Kame89,Hiya03,Hiya12FEW,Hiya12PTEP,Hiya12COLD} and by solving FY equation in configuration space \cite{LC_3n_PRC71_2005,LC_4n_PRC72_2005}.
 The complex energy solutions where computed using both the CS and ACCC methods. 
 The 3NF, 
 for both $T$ channels, were a sum of two hyper-radial Gaussians (one attractive and one repulsive):
\begin{equation}\label{V3NT}
V_{ijk}^{3N}=\!\!\sum_{T=1/2}^{3/2}\: \sum_{n=1}^2 W_n(T)e^{-(r_{ij}^2+r_{jk}^2+r_{ki}^2)/b_n^2} \, {\cal P}_{ijk}({T})
\end{equation}
 where ${\cal P}_{ijk}({T})$ is a projection operator for the total three-nucleon $T$ state.
 The parameters ($W_n,b_n$) for the $T=1/2$ channel were already fixed to reproduce the binding energies of some selected light nuclei and $^4$He$(e,e')^4$He$^*$ electromagnetic transition form factors \cite{Hiya04SECOND}, in conjunction with the AV8$'$ version of the Argonne NN potential \cite{AV8P97}.  
 For the $T=3/2$ channel, the $T=1/2$ parameters were used except for the attractive strength $W_1(3/2)$, the only parameter adjusted to reproduce RIKEN results.   

\begin{figure}[t]
\begin{center}
 \psfig{file=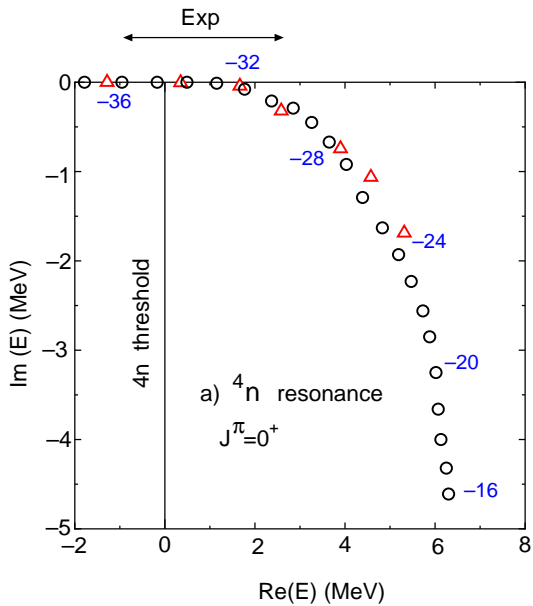,width=7.cm}\hspace*{5mm}
\caption{$^4$n resonance trajectory for the $J^\pi=0^+$ state. 
Circles correspond to AV8$'$ potential and triangles to INOY04 \cite{Dolesch}. The 3NF strength $W_1(3/2)$ was changed from $-37$ to $-16$~MeV  in steps of 1~MeV for AV8$'$, and from $-36$ to $-24$~MeV in steps of 2~MeV for INOY04. The RIKEN resonance region is indicated by the arrow at the top.}\label{4n_0+_Trajectory}
\end{center}
\end{figure}

 The critical value of $W_1(3/2)$ at which $4n$ becomes bound was determined for $J^{\pi}=0^+$, $2^+$, $1^+$, $0^-$, $1^-$... (by order of appearance). By decreasing this value, all of them moved into the resonance region, i.e.\ into the second Riemann sheet of the energy surface with $E=E_R-i E_I$.
 It was found however that the strength required to reproduce RIKEN results for the most favorable case ($0^+$) was $W_1(3/2)\in [-36,-30]$~MeV, which is totally unphysical. To have an idea, we can compare it with the corresponding $W_1(1/2)\approx 2$~MeV, while $T=3/2$ should be smaller \cite{AV18_IL2_2001} and subdominant in an EFT expansion \cite{Vnnn_2015}.
 Taking into account that $W_1(1/2)$ is responsible for about 5~MeV of extra binding in $^4$He, the additional energy required for the $4n$ system to become a resonant state is of the order of 100~MeV.
 Furthermore, it was found that starting from $W_1(3/2)\approx -18$~MeV the well-known unbound states $^4$Li, $^4$He$^*$ and $^4$H become bound.

 The trajectory of the $^4$n$(0^+)$ resonant state in the complex-energy plane as a function of $W_1(3/2)$ is shown in Fig.~\ref{4n_0+_Trajectory}. 
 Already at half the strength required to reproduce RIKEN result, the real part of the resonance is $E_R\approx6$~MeV and its width $\Gamma\approx 10$~MeV.
 But even these values are unphysically large, and they completely miss the description of the well-known $n$+$^3$H cross-section \cite{HLCK_4n_PRC93_2016}.
 The $3n$ was also studied in detail, and the conclusions were that the corresponding resonant states were located even further from the physical region than the $4n$ ones.

 In 2017 the same authors completed the previous work \cite{CHLK_4n_FBS_2017,LKH_PTEP_2017} by computing in their numerical simulations the strength function used by RIKEN experimental group, following their suggestion.
 While the works of 2005 \cite{LC_3n_PRC71_2005,LC_4n_PRC72_2005} had located the pole position of the nearest $^3$n and $^4$n resonances generated by realistic interactions, the question asked in 2017 was whether or not one could account for a $4n$ near-threshold resonance by modifying the most unknown part of the nuclear Hamiltonian: the 3NF for $T=3/2$.  
 Despite the different motivations and methods, all the conclusions were in line with the pre-GANIL era works: any near-threshold observable resonance in the $^3n$ and $^4n$ system (and a fortiori a bound state) must be definitely excluded.
 Concerning the RIKEN result, it was pointed out that there exist quantum mechanical mechanisms, other than $S$-matrix poles, that could generate a low-energy enhancement in a scattering cross-section.

 All in all, it seemed clear at this stage that Pieper's prediction of a few MeV tetraneutron resonance should be considered as an isolated case in the theoretical landscape, maybe a consequence of the trap artifact or of his cavalier extrapolation to the continuum, but in any event devoid of any physical ground. However, if this was clear for some people, it was not clear for everybody.

 Indeed, in the same 2016 Shirokov \etal\ \cite{Shirokov:2016ywq} found a $^4$n resonance with parameters $E_R=0.84$~MeV and $\Gamma=1.38$~MeV, in perfect agreement with RIKEN results.
 They used their non-local NN interaction JISP16 \cite{JISP16_2007} and  attempted several methods to describe the $^4$n state predicted by Pieper \cite{Pieper:2003dc}. The JISP16 $V_{NN}$ is expanded in terms of HO radial basis $R_{nl}$ as:
\[ V_{LL'SJ}(r,r')= \sum_{n=0}^{n(L)}    \sum_{n'=0}^{n'(L)}  \; R_{nL} (r)  \;    C_{nn'}^{LL'SJ} \; R_{nL'} (r')    \]
 with $n(L)=n'(L)=4$. It is thus a rank-4 separable interaction, whose $C_{nn'}$ coefficients and $b$ parameter of the HO were adjusted to successfully reproduce the low-energy properties of NN and some light nuclei. It can be considered as a high-precision potential.

 First they tried to solve the $4n$ problem with the No-Core Shell Model (NCSM) by scaling the interaction and tracking the lowest state as a function of the scaling factor $s$. They performed extrapolations to the unbound regime similarly to Pieper \cite{Pieper:2003dc} and found results in quantitative agreement, i.e.\ a resonance at $E_R\sim2$~MeV without any information on the width.
 Then they used No-Core Gamow Shell Model (NCGSM) techniques, which directly provide the resonant parameters in the complex energy plane.
 In this way they obtained a relatively broad resonance at $E_R\sim2.5$--3~MeV and $\Gamma\sim2.5$--6~MeV.
 They noted however that the resonance parameters decreased when the size of the NCGSM basis increased.

 Finally they used extended NCSM techniques to deal with the $4n$ in the continuum. In particular, in order to calculate  the $4\to4$ $S$-matrix resonant parameters, they used a Single-State Harmonic Oscillator Representation of Scattering Equations (SS-HORSE), an extension of the NCSM Hamiltonian in an HO basis by an infinite kinetic energy matrix.
 For this extension they used the so-called ``democratic decay approximation'', which describes the $4n$ continuum in terms of an  hyper-spherical harmonics (HH) basis, limited in practice to its ``minimal approximation'', i.e.\ with hyper-spherical momentum $K=2$.

 The SS-HORSE approach provides the $4\to4$ scattering matrix $S(E)$, and the corresponding phase shifts $\delta(E)$, at the positive eigen-energies of the NCSM Hamiltonian, from which they extrapolated the position of the $S$-matrix poles following some analytic expressions.
 These results depend critically in the parametrization of $S(E)$ in terms of four adjusted parameters.
 The values so obtained for the resonance parameters of the $S$-matrix pole were considered ``surprisingly small''.
 Their final result, almost exactly the same than RIKEN one, was obtained by further adding a false pole plus an additional fitting parameter, and was presented as a ``prediction''. 

 The findings of Shirokov, in line with Pieper's, were in strong contradiction with all the previous ones obtained with direct FY and Gaussian methods, and brought even more confusion to an already quite confuse situation.

 In 2017 Gandolfi \etal\ \cite{Gandolfi:2016bth} used Pieper's methods (Woods-Saxon trap, QMC techniques and extrapolation to the continuum) and reached a similar result: a $4n$ near-threshold resonance at $E_R=2.1(2)$~MeV. 
 Furthermore, they concluded that a $3n$ resonance was ``lower in energy than the tetraneutron one'', at $E_R=1.1(2)$~MeV.
 This was in fact the main question raised in the paper title, and with their positive answer the authors encouraged further experimental measurements.
 The resonance widths could not be estimated,
 but this work improved in several aspects the previous QMC calculations: the accuracy of the computed binding energy, the development of a consistent NN and 3NF chiral EFT interaction at next-to-next-to-leading order (N2LO), and the introduction of a quadratic instead of a linear extrapolation to reach the continuum.
 However, as we already noted and will develop below, this approach suffers from two inherent redhibitory drawbacks: the use of a 
 trap itself and the procedure of extrapolation to the continuum.

 One interesting conclusion was that, due to the extreme diluteness of the system, the roles played by the 3NF and the details of the NN one were very small.
 As a check of their trap results, they repeated the calculation by introducing an overall scaling factor $s$ on the $nn$ interaction. Surprisingly, they found that a value of only $s=1.3$ was enough to bind the $4n$, in sharp contradiction with all the previous works where an $s$ value 2--3 times bigger had been found.
 Pieper had also concluded that changes more dramatic than 30\% were needed.
 Even so, the extrapolated result at the physical value $s=1$ was  $E=2.0(1.0)$~MeV, consistent with their trap results.

 Also in 2017, Fossez \etal\ \cite{Marek_4n_2017} used two {\it ab initio} techniques and several NN interactions (chiral EFT and JISP16) to study the existence of a $4n$ resonance, which ``would deeply impact our understanding of nuclear matter''.
 The first technique was the NCGSM (as in Ref.~\cite{Shirokov:2016ywq}), which is a NCSM extension into the complex-energy plane with couplings to the continuum by means of the Berggren basis, that incorporates some selected resonances.
 The second technique was the Density Matrix Renormalization Group (DMRG), an alternative way to solve the nuclear many-body problem in the continuum.
 Both methods were supplemented by the use of natural orbitals and a new identification technique for broad resonances.

\begin{figure}[t] 
\begin{center}
  \psfig{file=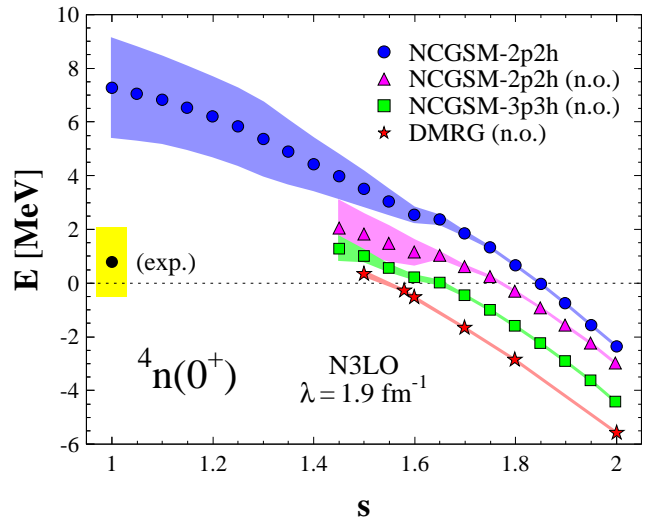,width=8.5cm}
 \end{center} \caption{Evolution of the energy and width (shaded area) of the $4n$ system with the scaling of the N3LO interaction from 2 to 1. The circles represent the NCGSM results with two neutrons in the continuum, which is used to generate the NCGSM results based on natural orbitals with two (triangles) and three (squares) neutrons in the continuum.
 The DMRG results without truncations are represented by stars. The RIKEN experimental energy is indicated by the solid circle, and the yellow area shows the uncertainty. Adapted from Ref.~\cite{Marek_4n_2017}.} \label{Fig_Marek}       
\end{figure}

 The $nn$ interaction was scaled by a global factor $s$ in order to bind the $4n$ system, which was then decreased 
 until it reached the physical state $s=1$.
 The results depend strongly on the particular technique (NCGSM or DMRG) and the different approximations made (see Fig.~\ref{Fig_Marek}).
 For example, at $s=2$ the $4n$ system is bound in all the approximations, but its binding energy ranges from $\approx 2$--6~MeV. Only in the less accurate calculation (NCGSM-2p2h) the physical value could be reached, with $E_R\approx7$~MeV and $\Gamma\approx4$~MeV.
 The other calculations could not go below $s\approx1.5$.
 They observed a clear trend in growing width as the $4n$ continuum was more accurately described.
 They even noted that ``the opening of new decay channels and the presence of continuum states in the configuration mixing above the threshold is expected to make the width explode when $s\to1$, especially in the DMRG results where all decay channels are open''.

 In view of this dispersion it is very difficult for the reader to draw a conclusion. In any event, the authors confirmed the existence of an $S$-matrix pole of the $4n$ system associated with $J^\pi=0^+$.
 They concluded that while the energy position of a $4n$ resonance might be compatible with the RIKEN value when $s\to 1$, including more than two particles in the continuum suggested that the width would be larger than $\Gamma \approx4$~MeV.

 Although sharing some methodological aspects, this work is much more careful in the analysis than Shirokov's, and is free from daring extrapolations. 
 The only objections we can think of, 
 besides the limited model space and the dependence on the HO parameter mentioned by the authors,
 is the use of a global scaling factor $s$ and the anomalous transition from an artificially bound state into the continuum that it generates.
 The appearance of $^2$n and $^3$n open channels can substantially modify the properties of the computed resonance when the system goes into the continuum.
 One may find intriguing that in all the considered schemes the critical scaling factor 
 determining the $4n$ threshold lies in the range $s\sim 1.5$--1.9, sensibly smaller than in most previous works (except  Gandolfi's $s=1.3$ \cite{Gandolfi:2016bth}).
 It is also unclear whether the Berggren basis used to extend NCSM into the continuum is adapted to describe a broad resonance, or if the results 
 are conditioned by the particularities of the Berggren basis itself. 

 In 2018 Deltuva studied $3n$ \cite{AD_3n_PRC97_2018} and $4n$ \cite{AD_4n_PLB782_2018} states in the framework of AGS equations in momentum space \cite{AGS_1967}.
 This formalism, equivalent to FY equations, provides rigorous solutions for the three- and four-body problem and describes very accurately all the corresponding nuclear systems around the breakup thresholds \cite{FD_FBS58_20917,DF_PRL113_2014,FCD_JPG41_2014}. 
 An important advantage of the AGS approach with respect to previous calculations is its ability to compute the resonance position and its effect on the scattering amplitudes that could lead to observables in a scattering process, in particular the $3n\to3n$ and $4n\to4n$ transition amplitudes.

\begin{figure}[t] 
\begin{center}
  \psfig{file=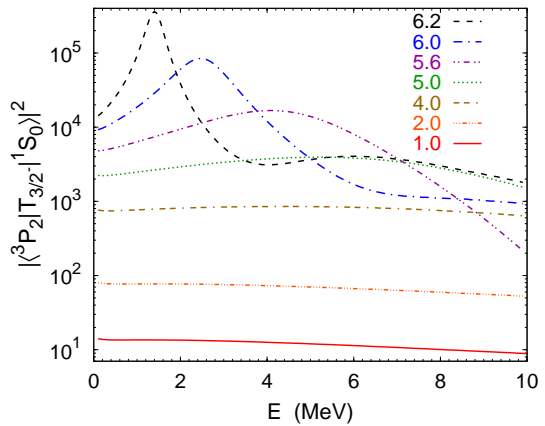,width=7.cm}
 \end{center} \caption{$3n$ transition amplitude for different values of the scaling factor in $^3$PF$_2$ partial wave, taken from Ref.~\cite{AD_3n_PRC97_2018}.} \label{3n_TA}       
\end{figure}

\begin{figure}[h!] 
\begin{center}
  \psfig{file=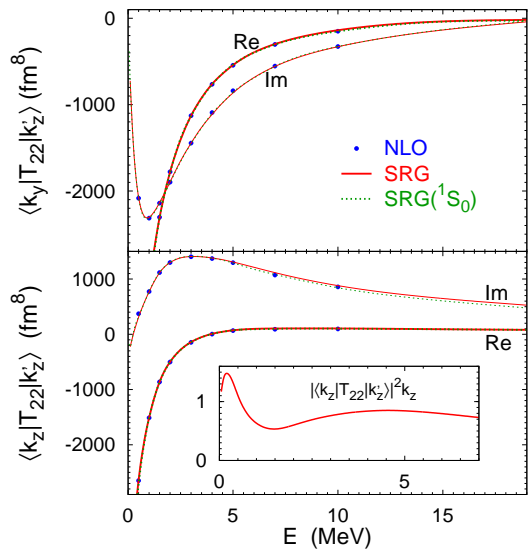,width=7.cm}
 \end{center} \caption{Physical $4n$ transition amplitude as a function of the energy, taken from Ref.~\cite{AD_4n_PLB782_2018}.} \label{4n_TA}       
\end{figure}

 The trineutron \cite{AD_3n_PRC97_2018} was studied using several realistic $nn$ interactions and with a scaling factor $s_{\alpha}$ only in the non-zero angular momentum partial waves, in order to avoid a bound $^2$n in $^1$S$_0$ that would open a threshold in the $3n$ system and complicate the analysis.
 For sufficiently large values of $s_{\alpha}$, a $^3$n bound state was found for several $J^{\pi}$ and at critical values within $s\approx6$--8, depending on the $nn$ interaction model.
 When decreasing $s_{\alpha}$, a resonance behavior was observed in all the transition amplitudes, as shown in Fig.~\ref{3n_TA} for the $^3$PF$_2$ case.
 Clearly, the resonant behavior for $s=6.2$ weakens when decreasing $s$ and completely disappears for the physical interaction strength.
 Note that even with $s=4$ the amplitude is totally flat.
 The corresponding pole trajectories 
 were also extracted, and for the Reid93 interaction they matched the results from FY calculations \cite{LC_3n_PRC71_2005}.
 Deltuva concluded that there are no physically observable $3n$ states consistent with the presently accepted interaction models,
 and that the pole trajectories were similar for all the realistic $nn$ potentials and in close agreement with Refs.~\cite{LC_3n_PRC71_2005,HLCK_4n_PRC93_2016}.

 The tetraneutron \cite{AD_4n_PLB782_2018} was studied using the same rigorous methodology.
 The $4n$ transition operators $T_{\beta\alpha}$ in the $0^+$ state were calculated using a physical $nn$ potential, i.e.\ $s_{\alpha}=1$. 
 As in the $3n$ case it showed no indication of resonances, but for sufficiently large $s_{\alpha}$ a resonant behavior was clearly seen in all matrix elements. The tetraneutron became bound at $s=5.29$, thus lower in energy than the trineutron, bound only at $s>6$.
 The results were found to be almost independent of the particular $nn$  potential, as in previous works.
 Interestingly, it was found that even in the absence of any $4n$ resonance, some $4\to4$ transition operators exhibit a low-energy enhancement (Fig.~\ref{4n_TA}, bottom insert).
 It was conjectured that they could also manifest in other reactions with the $4n$ subsystem in the final state, {like the $^4$He($^8$He$,^8$Be) one measured at RIKEN.}

 The conclusions of these two papers were in close agreement with the 
 exact FY methods \cite{LC_3n_PRC71_2005,LC_4n_PRC72_2005,HLCK_4n_PRC93_2016,SRV_JPG23_1997,WG_3n_PRC60_1999}, but in sharp contradiction with Refs.~\cite{Pieper:2003dc,Shirokov:2016ywq,Marek_4n_2017}, and even with Ref.~\cite{Gandolfi:2016bth} that had used the same EFT potential.

 In 2019 Deltuva {and Lazauskas} shed some light on this striking difference 
 \cite{Comment_PRL123_2019,Arnas_Rimas_4n_2n_PRC100_2019}.
 The first paper was a critical Comment \cite{Comment_PRL123_2019} to the conclusions of Ref.~\cite{Gandolfi:2016bth}. 
 In particular, they analyzed quantitatively some serious shortcomings related to confining neutrons in a Woods-Saxon trap (\ref{V_T}) and to the misleading extrapolation procedure without taking into account the analytical structure of the thresholds.
 Two problems were pointed out:
\begin{enumerate}
\item
 For some trap parameters \cite{Gandolfi:2016bth}, e.g.\ $R=6$~fm and $a=0.65$~fm, $^2$n starts being bound at $V_0\sim0.09$~MeV.
Thus, when $^4$n is supposed to become bound in the trap, $^2$n is already bound by several MeV (see Appendix for details) and $^4$n should in fact decay into $^2$n(s) and maybe even $^3$n, provided the theoretical dynamical method was able to account for open scattering channels.
 In any case, the $4n$ states of Refs.~\cite{Pieper:2003dc,Gandolfi:2016bth} fulfilling $E_{2n}<E_{4n}$ were not really bound states, but some discretized continuum states that do not evolve into a resonance. The extrapolation of their energies into the continuum is misleading
\footnote
{It is indeed very disappointing that this trivial kinematical fact could have been ignored not only by the authors but also passed through the referee procedure of top level journals.}.
\item
 Even assuming that the $4n$ state in the trap was really bound, the evolution into a continuum state involves branching at each threshold with discontinuity in the second derivative of energy with respect to a strength parameter. Polynomial extrapolations neglect this discontinuity and are thus conceptually incorrect. The non-trivial evolution of the simple $2n$ case was considered in detail, and is reproduced in our Appendix.
\end{enumerate}
 Gandolfi \etal\ argued \cite{Reply_Comment_PRL123_2019} that they had not claimed the existence of $3n$ or $4n$ resonances, nor seen evidence of open decay channels in the neutron trap, and that their extrapolation had worked in some selected examples.




 In the second paper \cite{Arnas_Rimas_4n_2n_PRC100_2019} 
 Deltuva {and Lazauskas} studied the consequences of a bound $^2$n, obtained by enhancing $V_{nn}$ with a scaling factor, in the evolution towards the continuum of an artificially bound $^4$n state.
 The results were obtained by combining the FY \cite{LC_3n_PRC71_2005,LC_4n_PRC72_2005} and AGS \cite{AD_4n_PLB782_2018} formalisms. Two different cases were considered:
\begin{enumerate}
\item
 The same scaling factor $s$ was used in all partial waves. 
 They found that a $0^+$ bound $^4$n indeed emerges for a deep enough $^2$n, with $s\approx 2.4$--2.7 depending on the particular potential.
 When slightly decreasing $s$, this artificially bound $^4$n, although having negative energy, is not longer bound but decays into two $^2$n,
 in contradiction with Refs.~\cite{Gandolfi:2016bth,Marek_4n_2017} that found a bound $^4$n for $s\approx 1.5$.
 Furthermore, by continuing to decrease $s$, they found that the $4n$ system will never evolve into a resonance but into a virtual state,  similar to the NN  $^1$S$_0$ case.
 The binding is then dominated by the S-wave and it is clear that there is no way to build a $4n$ resonance with a global scaling factor.
\item
 The scaling factor for the $^1$S$_0$ interaction was fixed to $s=1.3$ (with $B_{nn}=0.316$~MeV) and the main extra binding was provided by scaling P- and higher angular momentum partial waves.
 This breaking of S-wave dominance is essential to produce P-wave resonances in the final state.
 Such a resonance is indeed formed with $s_P\approx3.9$, but when slowly removing the scaling factor the resonant character disappeared well before reaching the physical case $s=1$.
\end{enumerate}
Their main result, the fact that one scaling factor cannot generate a $4n$ resonance, is enlightening and very relevant in view of the many studies that use this artifact.

 Later in 2019 Li \etal\ \cite{Li_Michel_3n_PRC100_2019} contradicted the two previous works. They used {\it ab initio} NCGSM and concluded that $^3$n and $^4$n are both observable resonant states.
 The $^4$n resonance parameters were $(E,\Gamma)_{4n}=(2.64,2.38)$~MeV. They even claimed that the $^3$n was lower in energy and narrower, $(E,\Gamma)_{3n}=(1.29,0.91)$~MeV, and strongly encouraged experimentalists to search for the trineutron at low energy.
 These calculations, that pretend to be ``nearly exact'', are however based in the same trap methods of Refs.~\cite{Pieper:2003dc,Gandolfi:2016bth}, and 
we have already discussed the kind of redhibitory problems from which these methods suffer.

 The multineutron field is still very active, as shown by the two additional works devoted to this topic that have been already published in 2020, and with which we will close this theoretical review.
 
 In the first one, Higgins \etal\ \cite{HGKV_PRL125_2020} 
 studied the $3n$ and $4n$ systems within the adiabatic hyper-spherical framework. The corresponding adiabatic potential-energy curve was analyzed and found to be repulsive, in agreement with Timofeyuk \cite{Timofeyuk:2003ya}.
 They concluded that there is no sign of a low-energy resonance for any of these systems. However, they observed in both of them some low-energy enhancement of the Wigner-Smith ``time delay'', which for a single adiabatic potential is defined as 
 $ Q(E)\equiv { d\,[\delta(E)] / dE} $,
 that could provide a hint to understand the GANIL and RIKEN near-threshold enhancements.

 Later in 2020 Ishikawa \cite{Ishikawa_3n_PRC102_2020}, using FY 
 and realistic NN potential (AV18), found no evidence of a $3n$ resonance.
 Moreover, he examined different methods used to extrapolate the $3n$ energy from an artificially bound state to the continuum, and found that the enhancement of $nn$ P-wave alone or the use of 3NF do not generate spurious $3n$ resonances.
 However, using external trapping potentials leads to positive $3n$ energy results, which without a further careful analysis may be considered as a resonance while they are in fact a general defect of the trapping method.
 In this way, Ishikawa explained the contradictory results of Refs.~\cite{Gandolfi:2016bth,Li_Michel_3n_PRC100_2019} concerning $3n$ states.

\subsection{Conclusions from theory}

 At first glance, the striking disagreement among different theoretical works presented in this section is obvious, shocking, and could be even devastating for the credibility of the field. From a general point of view one can distinguish two main families of results, whose conclusions are orthogonal to each other.

 In the first group, the existence of near-threshold $3n$ and $4n$ resonant states is claimed. In the case of the tetraneutron, those works seem thus in agreement with the two experimental signals reported at GANIL and RIKEN.
 However, for the second group this is a scientific nonsense since an explicit calculation of these states finds them very far from the physical region, in the third quadrant of the unphysical energy sheet. 
 According to them, the recent experimental results, if confirmed, in no way can be attributed to a $3n$ or $4n$ resonant state and must have another origin, still to be clarified.

 If these two points of view can hardly be more opposite, there is at least a general consensus in that the origin of such contradictory conclusions must not be found in the different interactions used.
 Indeed, for these low-energy physical states with a large spatial extension, any interaction providing acceptable low-energy parameters  leads to similar qualitative results.
 Furthermore, the role of three-neutron forces is negligible.
 The differences among them must rather be found in the methods used to solve the few-nucleon problem and/or in the way they access the few-neutron continuum.
 
 All results obtained using FY or the AGS equations agree with each other about the non-existence of any near-threshold $^3$n and $^4$n.
 These approaches represent nowadays the most rigorous framework for solving the few-body problem, and include in a natural way all the continuum states, 
 without additional approximations or ansatzs.
 Their agreement is not only qualitative but also quantitative, in particular when computing the critical values of the scaling factors or the trajectories of the resonance position in the complex-energy surface.
 Moreover, this agreement is extended to the variational Gaussian Expansion methods, when the resonant states are tackled with CSM or ACCCM, and to HH, 
 where they unanimously find a strong repulsion in the adiabatic potential curves disallowing any bound or resonant state.

 Furthermore, we would like to mention the EFT developments in multifermion systems close to the unitary limit, as it is the case of multineutron states, where they found a universal dimer-dimer and fermion-dimer strong repulsion \cite{EFT_Petrov_2004,EFT_Rittenheuse_2011,EFT_dd_PLB_2017}, in contrast with 
 any resonant state and in line with the above mentioned series of works.

 This impressive bulk of coherent results is in frank contradiction with the two GFMC results, as well as with the NCSM-SS-HORSE and the two NCGSM works, the latter three being based on many-body techniques.

 GFMC \cite{Pieper:2003dc,Gandolfi:2016bth} is accurate for computing bound states and has been able to deal with a larger number of particles than FY and AGS, but it is clearly less adapted to scattering problems, in particular the computation of resonances.
 Moreover, we have shown that their conclusions are strongly weakened due to two important additional issues: the use of traps and the method employed to extrapolate the (supposed) bound state into the continuum.

 NCSM-SS-HORSE \cite{Shirokov:2016ywq} is based on several many-body techniques 
 that were not checked in similar few-nucleon scattering problems. 
 Moreover, the $4n$ continuum is treated in a very indirect way, 
 with the ``democratic decay'' hypothesis 
 for extracting the $4n\to4n$ amplitude (neglecting the FSI between virtual dineutrons \cite{Arnas_Rimas_4n_2n_PRC100_2019}) and a non-trivial extrapolation procedure to extract the complex parameters of an eventual resonance.
 Finally, they needed the further addition of a ``false pole'' in order to be in perfect agreement with the previously published RIKEN result.

 The more comprehensive NCGSM-DMRG work \cite{Marek_4n_2017} exhibits a manifest dependence in the approximations used to solve the four-body problem, and their global scaling factor 
 could reach the physical case only for the less accurate of them.
 However, they observe a clear trend of an increasing width of the state, which according to the authors ``exploded when $s\to1$'', and concluded with a resonant state that would be much larger than the reported RIKEN limit.
 We believe that the use of a global scaling factor is a major problem, which due to the dineutron formation could never generate a resonance but a virtual state \cite{Arnas_Rimas_4n_2n_PRC100_2019}.
 The most recent NCGSM work \cite{Li_Michel_3n_PRC100_2019} suffers from most of the previous drawbacks plus the use of a trap.

 Finally, any method that does not take properly into account the thresholds can lead to uncontrolled conclusions, and we believe that this is a common problem of the references in this second group.

 With respect to the experimental results, we cannot find any theoretical explanation for GANIL and RIKEN events, both corresponding to an enhancement of low-energy $4n$ in a given reaction cross-section, either breakup or DCX.
 We have presented in this section convincing arguments against the attribution of this enhancement to a resonance, at least in the standard way a resonance is understood in nuclear, hadronic and particle physics.

\begin{figure}[t]
\begin{center}
\psfig{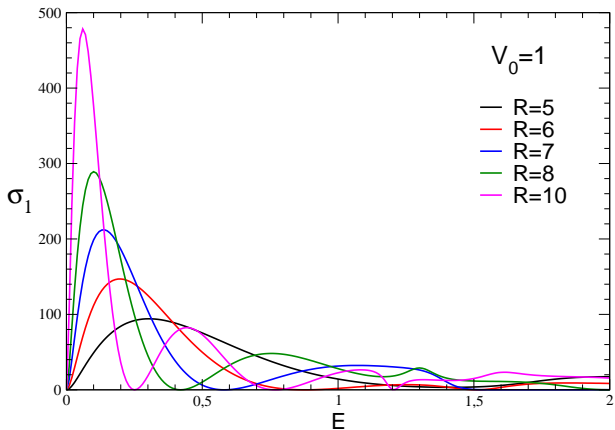}
\caption{Resonant P-wave cross-section in a repulsive potential of the form $V(r)=+V_0$ when $r<R$, for different values of $R$.}\label{Repulsive_Resonance}
\end{center}
\end{figure}

 However, as it has been emphasized many times and by many authors \cite{LC_4n_PRC72_2005,HLCK_4n_PRC93_2016,GTZ_EPJA19_2004,LKH_PTEP_2017,AD_4n_PLB782_2018,HGKV_PRL125_2020}, there are other quantum mechanical mechanisms that generate resonance-like structures, even for repulsive interactions. 
 \edit{Deltuva's results of Fig.~\ref{4n_TA} show it explicitly in the $4n\to4n$ non-resonant transition amplitude.}
 A much simpler case is illustrated in Fig.~\ref{Repulsive_Resonance}, 
 \edit{corresponding to the P-wave scattering of two identical particles ($\hbar^2/m=1$) on a repulsive potential well $V(r)=+V_0\times\Theta(R-r)$,
 where $\Theta$ is the Heaviside step function.
 By changing the parameter $R$, one can find a rich variety of resonance-like structures, with the corresponding $E$-dependent phase shift crossing $\pi/2$ close to the resonance peak\footnote{The same potential was considered in \cite{GTZ_EPJA19_2004} to illustrate an S-wave resonance-like behavior.}.}
 Ignoring the underlying dynamical content, these structures can be attributed to a resonant state, very close to binding.
 We believe that this kind of behavior is also manifest in the time-delay analysis of Ref.~\cite{HGKV_PRL125_2020}.

 Whether or not the experimental search of this quantum mechanical enhancements is pertinent enough to justify great investments can be submitted to debate.
 However, associating these enhancements to physical resonant states would be misleading.
 Waiting for the next theoretical paper claiming near-threshold $3n$ and $4n$ resonances, it seems clear that after all these decades of efforts and discussion a general consensus based on rigorous solutions of the problem is emerging to conclude... what was already clear from the first Gl\"ockle's work in 1978.

\section{Sit finis libris...}\label{Sec_Conclusions}

 ``Sit finis libris, non finis quaerendi'' (this may be the end of the
book, but not the end of the quest), was an elegant way to close an open
question in the scholastic times.

 Despite all the controversies and debates, this quest is definitely a
fascinating story. Think of the distant 1960s, with experimentalists
trying to do `magic' with pions in order to create neutron matter in the
laboratory; others going to nuclear reactors, looking for their grail in
such a radioactive environment; others manipulating ion beams, hoping to
add an remove nucleons against all odds.
 In parallel, theoreticians starting the first three-body calculations,
despite lacking at that time the sophisticated tools and inputs that we
have seen were needed.
 Sixty years later the quest is still fascinating... but still open.
 If from the theoretical point of view the question of a narrow $3n$ and $4n$ resonance seems closed, there is an urgent need for a definitive experimental conclusion concerning the GANIL and RIKEN events.

 Nuclear theory, and in particular the description of light nuclei from
the underlying forces between nucleons, has seen an enormous progress
working hand in hand with experiments. The main theoretical models have
been born from the known properties of stable nuclei, and then they have
been refined and updated with the increasing knowledge about more and
more exotic isotopes.
 In the field of neutral nuclei, however, the theory advances almost
blind.
 The only two experimental signals \cite{Marques02_4n_recoil,Kisamori16_4n_DCX} are still weak 
 and do not provide any firm and precise observable that could benchmark the different methods and techniques.

 In this respect, it should be a priority to confront the controversies surrounding the tri- and tetraneutron calculations with the help of high-statistics, unambiguous experimental results. New techniques or experiments that would provide still another weak and/or ambiguous signal do not seem the best way to unlock progress in the field.
 It is our hope that the new experiments already (or soon to be) undertaken at RIKEN will bring the very much needed reliable reference points for the theoretical conclusions to be adapted.
 If the positive signals were refuted, then the calculations predicting resonances should be reevaluated. If they were confirmed though, then the detailed characteristics of the signals \edit{(such as energies, widths, or angular correlations) should be explained.}

 Once the $3n$ and $4n$ systems will be clarified, both from theory and experiment, it will be easier to move on the heavier systems, in particular $6n$ and $8n$.
 Even if the lightest multineutron states were not observable, a pertinent question remains: when increasing the mass number, can at some point a multineutron manifest as a bound or resonant state? Even for the most reluctant theories, and of course for experiments, this is still an open question.
 The analogy between neutrons and atomic $^3$He is very suggestive, for their fermionic character and for the form of their interaction, with a hard repulsive core and a small attractive well.
 The dineutron (with a critical enhancement factor $s\approx 1.1$) is even closer to binding than the $^3$He dimer ($s\approx 1.3$), and we know that a $^3$He drop becomes bound beyond $N\sim30$ atoms \cite{Guardiola_PRL84_2001}.
 On the other hand, the $nn$ P-waves are smaller and its centrifugal barrier is less peripheral than in the helium atomic case \cite{LC_4n_PRC72_2005,Rimas_PhD_2003}, and that could make a difference.

  The known as ``helium anomaly'', i.e.\ the fact that $^8$He is more bound than $^6$He, has long been a clue towards the search for light multineutrons, taken as a hint for additional binding due to the increasing number of neutrons on top of $^4$He.
 Less known but maybe more spectacular would be the analog ``hydrogen anomaly'', provided it is established.
 The $^{4,5,6}$H isotopes are unbound by several MeV and exhibit broad resonances, but $^5$H ground state is narrower than the other two, as
demonstrated by recent {\it ab initio} calculations of these resonances \cite{LHC_PLB791_2019}.
 Moreover, $^7$H is reported to lie almost at the $4n$ threshold \cite{H7_GANIL_2007}, although it has in common with the tetraneutron that the experimental signals to date are weak, ambiguous and sometimes contradictory.

 With the recent progress in the exact calculation of the five-body problem, $^7$H is now accessible almost {\it ab initio} (treating the triton as a particle) for the theory and, as we have seen at the end of Sec.~2, a related experiment that should provide very high statistics and resolution has been already undertaken.
 This super-heavy isotope of hydrogen could in this way be the key for the next steps in the field, both for the role of the tetraneutron at the $t+4n$ threshold but even for the hexaneutron at the $p+6n$ threshold.
 Going beyond $A=6$ represents still a too important obstacle for experiments, and for the moment should be left for the theory, once the lighter multineutrons have helped to provide a solid base for the models.

 In any event this domain will remain fascinating since, paraphrasing
the authors of Ref.~ \cite{Marek_4n_2017}, it ``would deeply impact our
understanding of nuclear matter''. 


\appendix

\section{Two neutrons in a trap}\label{App1}

 In order to illustrate the kind of problems related to the extrapolation from bound states to the continuum, we present in this Appendix a detailed solution of the simplest case, \edit{in the same spirit as it was considered in Ref.~\cite{Comment_PRL123_2019}}: two neutrons confined in a trap.
 
 Let us consider two neutrons interacting via $V_{nn}$ and furthermore submitted to a potential trap having, as in Refs.~\cite{Pieper:2003dc,Gandolfi:2016bth,Li_Michel_3n_PRC100_2019}, a Woods-Saxon form:
\begin{equation}\label{V_Ti}
 V_T(r_i)=  {V_0\over  1 +e^{  {  r_i-R \over a }     }  }
\end{equation}
 The total two-body Hamiltonian is:
\begin{equation}\label{Hnn_T}
 H= -\frac{\hbar^2}{m_n}\Delta +V_{nn}(\vec{r}_2-\vec{r}_1) + V_T(r_1) + V_T(r_2)  
\end{equation}
 The particle coordinates $r_i$ can be expressed in terms of the relative ($\vec r=\vec r_2-\vec r_1$) and center-of-mass ($2\vec R=\vec r_1+\vec r_2$) coordinates as:
\begin{eqnarray*}
 \vec r_1-\vec r_2 = 2\vec r_1 - (\vec r_1+ \vec r_2) &\Rightarrow& \vec r_1= - {\vec r\over2} + \vec R   \cr
 \vec r_2-\vec r_1 = 2\vec r_2 - (\vec r_1+ \vec r_2) &\Rightarrow& \vec r_2 =+{\vec r\over2} + \vec R
\end{eqnarray*}
 Assuming $\vec{R}\equiv0$ to get rid of the center-of-mass motion, the intrinsic two-body Hamiltonian takes the form:
\begin{equation}\label{Hn_T}
 H=  -{\hbar^2\over m_n}\Delta +V_{nn}(r) + 2V_T\left( {r\over 2} \right) 
\end{equation}
 The numerical solution of this problem is trivial and can be obtained with any required accuracy.
 For illustrative purposes, we have considered a phenomenological CD MT13 $nn$ interaction:
\begin{equation}\label{CD_MT1} 
V_{nn}(r)=  V_r \frac{\exp(-\mu_r r)}{r} - V_a \frac{\exp(-\mu_a r)}{r}   
\end{equation}
 reproducing the $nn$ scattering length $a_{nn}=-18.59$~fm and effective range $r_{nn}=2.94$~fm values,
 with the parameters $V_r=1438.72$~MeV, $\mu_r=3.11$~fm, $V_a=509.40$~MeV, $\mu_a=1.55$~fm, and ${\hbar^2/ m_n}=41.44250$~Mev\,fm$^2$ \cite{HLMC_PRCRC_2019}.

\begin{figure}[t] 
\begin{center}
  \psfig{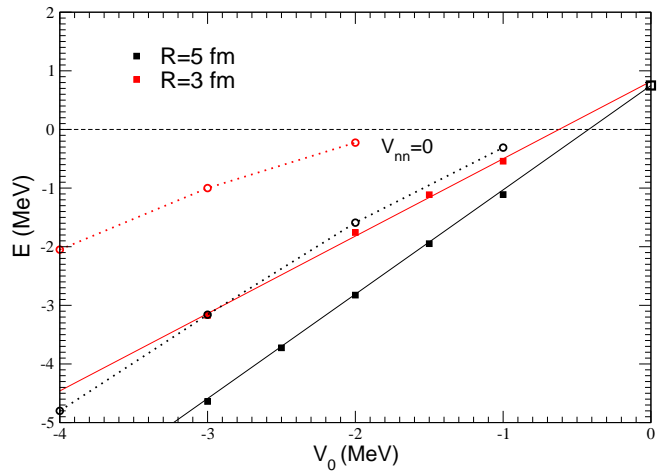}
 \end{center} 
 \caption{Energy of $2n$ in a trap (\ref{V_Ti}) as a function of the strength parameter $V_0$ for two values of the trap size $R$.
 A linear extrapolation (solid lines) converges towards a positive energy value. The free case $V_{nn}=0$ (dashed lines) displays a similar trend.}\label{FIG_n2_Trap_LINEAR}
\end{figure}

 The binding energies for two values of the trap size ($R=3$ and 5~fm) and fixed diffusiveness $a=0.65$~fm are plotted in Fig.~\ref{FIG_n2_Trap_LINEAR}
as a function of the strength $V_0$.

 At first glance it seems that both series of points 
 can be indeed linearly extrapolated (solid lines) towards a positive value (empty square at $V_0=0$), which could be interpreted, as in Refs.~\cite{Pieper:2003dc,Gandolfi:2016bth,Li_Michel_3n_PRC100_2019}, as a $2n$ resonance with positive energy.
 Furthermore, it is worth noting that even in the absence of any interaction, the results of the linear extrapolation (dashed lines) display a similar trend and would lead to a very disturbing resonant state between two non-interacting neutrons!

\begin{figure}[t]
\begin{center}
\psfig{file=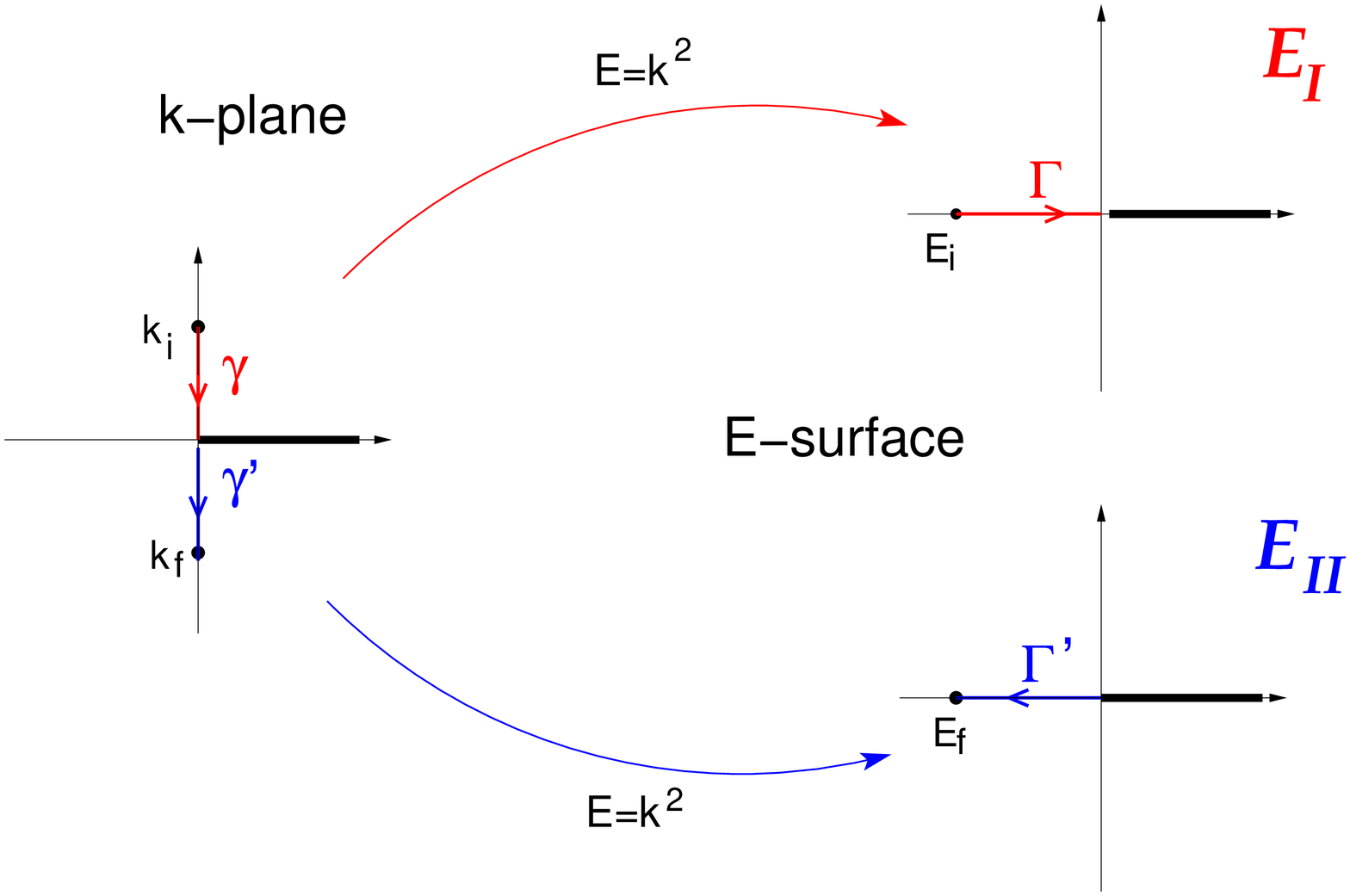,width=8.5cm}
\caption{Pole trajectory  in the complex-momentum plane ($\gamma\cup\gamma'$) and in the $E$-surface ($\Gamma\cup\Gamma'$) of an artificially bound $nn$ ($k_i$ and $E_i\in E_I$) evolving to the continuum ($k_f$ and $E_f\in E_{II}$).}\label{nn_K_E_Surface}
\end{center}
\end{figure}

 One could argue that in the case of two neutrons in a $^1$S$_0$ state, the extrapolated results of Fig.~\ref{FIG_n2_Trap_LINEAR} correspond in fact to the $nn$ virtual state with positive energy, as it is often considered. 
 We remind, however, that the virtual state, corresponding to a pole in the negative imaginary axis $k_0=+i\kappa$, has negative energy $k_0^2=-\kappa^2$.
 The pole trajectory in the complex $k$-plane and $E$-surface of an artificially bound $nn$ ($k_i$ and $E_i\in E_I$) into the continuum ($k_f$ and $E_f\in E_{II}$) is summarized in Fig.~\ref{nn_K_E_Surface}.
 As one can see, both bound and virtual (unbound) states correspond to negative-energy values in the $E$-surface, though living in different Riemann sheets.

\begin{figure}[t]
\begin{center}
\psfig{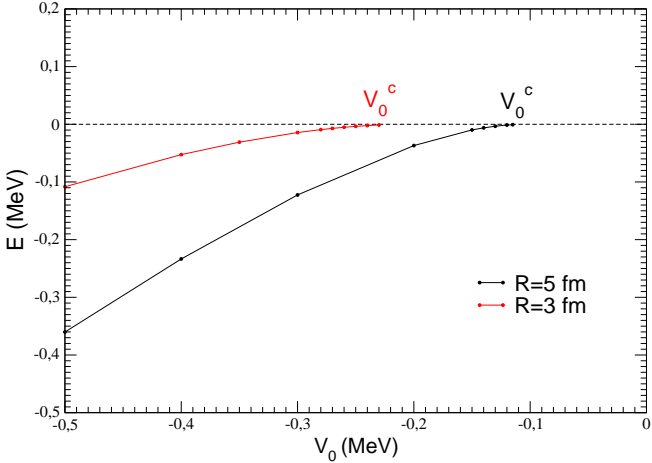}
\end{center}
\caption{Zoom of Fig.~\ref{FIG_n2_Trap_LINEAR} showing the behavior of the energy when approaching the continuum threshold at $V_0^c$ and $E=0$.}\label{FIG_Trap_Zoom}
\end{figure}

 If the extrapolation analysis sketched in Fig.~\ref{FIG_n2_Trap_LINEAR} is done carefully, in particular using much smaller binding energies, one can see that a strong curvature in the $E(V_0)$ dependence starts at $B\approx0.1$~MeV until the continuum threshold $E=0$ is reached at a critical value $V^c_0(R)$, which increases with $R$.
 For instance, $V^c_0\approx0.1$~MeV for $R=5$~fm (see the zoom displayed in Fig.~\ref{FIG_Trap_Zoom}).

\begin{figure}[t]
\begin{center}
\psfig{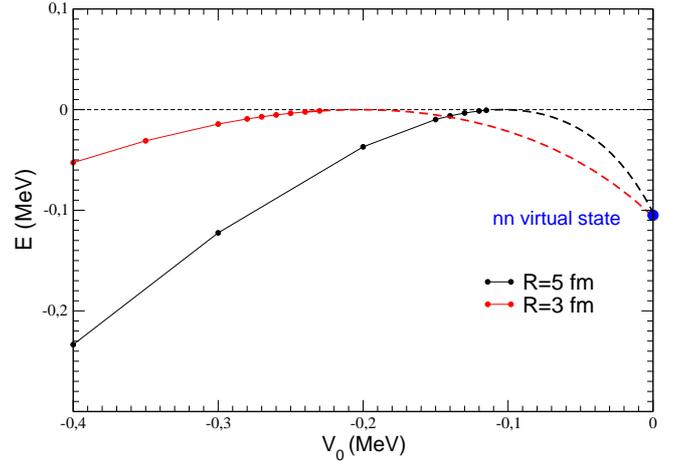}
\end{center}
\caption{Energy of $2n$ in a trap, from bound state to continuum. 
The corresponding trajectory in the complex-momentum plane ($\gamma\cup\gamma'$) and the $E$-surface ($\Gamma\cup\Gamma'$) is displayed in Fig.~\ref{nn_K_E_Surface}.}\label{Fig_n2_Trap}
\end{figure}

 After the threshold, the $V_0$-dependence represented in Fig.~\ref{FIG_Trap_Zoom} changes drastically the curvature, and turns downwards toward negative values of $E$ as can be see in Fig.~\ref{Fig_n2_Trap}, and as expected from the location of a virtual state in the continuum illustrated in Fig.~\ref{nn_K_E_Surface}.
 The highly non-trivial trajectory in the $(E,V_0)$ plane displayed in Fig.~\ref{Fig_n2_Trap}, results from the ACCC extrapolation given by equation (\ref{k_NM}).
 To be faithful, a set of bound state values $(V_{0}^n,E^n)$ must be computed with very high accuracy, as accurate should be the critical values of the potential strength at threshold $V_0^c$, when the system enters the continuum. 
 As an example of the accuracy required, the values used for the case $R=5$~fm are listed in Table~\ref{Table_E_Trap}.

 The trajectory in the $(E,V_0)$ plane is represented in the complex  $k$-plane in the left panel of Fig.~\ref{nn_K_E_Surface}. It is the straight line formed by the union of $\gamma\cup\{0\}\cup\gamma'$.
 The corresponding path in the $E$-surface is $\Gamma\cup\{0\}\cup\Gamma'$, shown in the right panel, where $E_i$ belongs to $E_I$ and $E_f$ to $E_{II}$. 
 The transition from bound state to continuum is at $k=0$ and coincides with the transition $E_I \to E_{II}$.
 Note that both initial ($k_i$) and final ($k_f$) states  have negative energy ($E_i$ and $E_f$ respectively), despite the fact that $E_f$ is in the continuum, but belong to different Riemann sheets.

\begin{table}
\begin{center}
\begin{tabular}{l l }
$V_0$    &  $E$  \\\hline   
0.20  &    -0.036938 \cr
0.15   &   -0.009817  \cr
0.14    &  -0.006220 \cr
0.13    &  -0.003367 \cr
0.12    &  -0.001331 \cr
0.115   & -0.00064676 \cr
0.10384(1)            &  0 
\end{tabular}
\caption{Energy values of $2n$ in a trap used in the ACCC extrapolation (\ref{k_NM}), as a function of the potential strength $V_0$ for $R=5$~fm. The last point corresponds to the threshold extrapolated value.}\label{Table_E_Trap}
\end{center}
\end{table}

 From the above results it is clear that a simple polynomial extrapolation cannot describe the non-trivial evolution of a bound state into the continuum displayed in Fig.~\ref{Fig_n2_Trap}.
 This is in fact the main conclusion of this Appendix.
 ACCCM can do it accurately by taking into account the analytical properties related to the thresholds. However, the method requires the previous computation of several bound state values with high accuracy and very close to $E=0$. 
 This can be easily done in the two-body case we have considered above, although it becomes more and more involved when increasing the number of particles.
 This level of accuracy is not accessible to all the methods used to solve the few-nucleon problem, but it has been successfully applied to $A=3,4,5$ in Ref.~\cite{LC_3n_PRC71_2005,LC_4n_PRC72_2005,HLCK_4n_PRC93_2016,LHC_PLB791_2019}.

\edit{The case presented above corresponds to an S-wave virtual state, which is different from the realistic $3n$ and $4n$ P-wave resonances displayed in Figs.~\ref{3n_Gloeckle_32m},\ref{FIG_TRAJEC_3n_4n}.
 They have however in common that the trajectories of the corresponding pole positions end at negative energy, \mbox{Re$(E)<0$}, in the third quadrant of the second Riemann sheet $E_{II}$, as illustrated in Fig.~\ref{nn_K_E_Surface}.
 Of course, it is clear that a polynomial extrapolation of the results displayed in Fig.~\ref{FIG_n2_Trap_LINEAR} (or Fig.~\ref{E4n_WS_Trap}) is forced to end at positive energy values, and is therefore unable to reproduce the realistic behavior of the $3n$ and $4n$ resonant states.}

\begin{figure}[t] 
\begin{center}
  \psfig{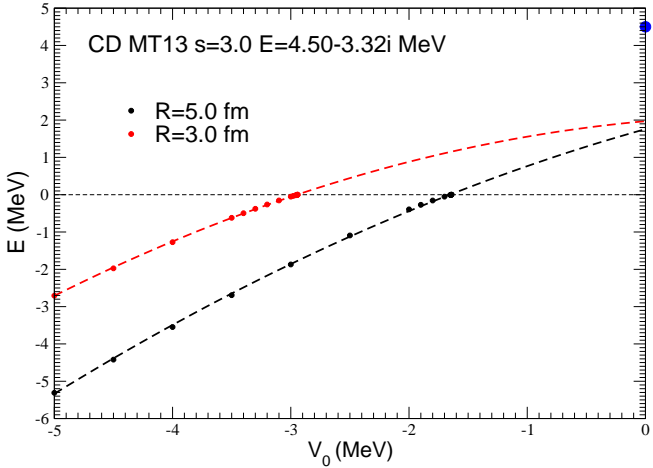}
 \end{center} 
 \caption{\edit{Energy of $2n$ P-wave resonant state in a trap (\ref{V_Ti}) as a function of the strength parameter $V_0$ for two values of the trap size $R$.
 Quadratic extrapolations (dashed lines) of the computed binding energies in the trap converge at $V_0=0$ towards positive energy values, sensibly different from the exact result (blue circle)}.}\label{FIG_P_nn_Trap}
\end{figure}

\edit{Furthermore, it is worth noting that even in the case of positive energy P-wave resonances, with Re$(E)>0$ and Im$(E)<0$ like the one represented in Fig.~\ref{4n_0+_Trajectory}, a polynomial extrapolation disregarding the analytic threshold structure can lead to wrong results.
 This can be illustrated by considering the same $nn$ interaction (\ref{CD_MT1}) but suitably scaled in order to generate a P-wave $nn$ resonant state. For instance, with a scaling factor $s=3.0$ one has a resonance at $E=4.40-3.32\,i$~MeV, that is a state with $\Gamma\approx6.6$~MeV, still smaller than the realistic case of  Fig.~\ref{4n_0+_Trajectory}.
 To complete our results for $nn$ S-wave, we have computed the energies of this P-wave resonant state in the trap, as a function of $V_0$ for two different parameters of the trap size $R$, and performed 
a quadratic extrapolation towards the `free' case $V_0=0$.
 Results are displayed in Fig.~\ref{FIG_P_nn_Trap}. They show a significant disagreement between the extrapolated values and the physical resonance energy, indicated by a blue solid circle.
 We cannot exclude, as noted in Ref.~\cite{Comment_PRL123_2019}, that for very narrow P-wave resonances the difference could be small, but 
the polynomial extrapolation is not reliable in a general case.
 This could be the case in the unphysical example (P-wave resonance in an S-wave two-Gaussian potential) considered in Ref.~\cite{Gandolfi:2016bth} in order to justify their extrapolation method.}

 \edit{The exact solution of two neutrons in a trap} can be straightforwardly extended to the $^3$n case.
 The Jacobi coordinates of the $3n$ system are expressed in terms of the particle coordinates $r_i$ as:
\begin{eqnarray*}
 \vec{y}_1 & = & \frac{2}{\sqrt{3}} \left( \vec{r}_1 - \frac{\vec{r}_2+\vec{r}_3}{2} \right) \cr
  & = & \frac{2}{\sqrt{3}} \left( \vec{r}_1 + \frac{\vec{r}_1}{2} - \frac{\vec{r}_1+\vec{r}_2+\vec{r}_3}{2} \right)
\end{eqnarray*}
 Since in the center of mass $\vec{r}_1+\vec{r}_2+\vec{r}_3=0$, one has:
\[ \vec{r}_1= \frac{\vec{y}_1}{\sqrt{3}}  \]
 and equivalent expressions for the other $r_i$.

 The three-body intrinsic Hamiltonian takes the form:
\[ H=H_0+V_{nn}(x_1)+V_{nn}(x_2)+V_{nn}(x_3)+W(\vec{y}_1,\vec{y}_2,\vec{y}_3) \]
 with:
\[ W(\vec{y}_1,\vec{y}_2,\vec{y}_3)=V_T\left(\frac{\vec{y}_1}{\sqrt{3}}\right)
 +V_T\left(\frac{\vec{y}_2}{\sqrt{3}}\right)+V_T\left(\frac{\vec{y}_3}{\sqrt{3}}\right) \]

 Since the different sets of Jacobi coordinates are related by the linear transformations:
\begin{eqnarray*}
 \vec{y}_2 &=& c_{21} \vec{x}_1 + s_{21} \vec{y}_1 \cr
 \vec{y}_3 &=& c_{31} \vec{x}_1 + s_{31} \vec{y}_1 
\end{eqnarray*}
 the problem is equivalent to adding a three-body force:
\[ W(x_1,y_1) =   W[ \vec y_1,\vec y_2(\vec  x_1, \vec y_1) ,\vec y_3(\vec x_1,\vec y_1)] \]
 which in the case of three identical particles can be easily solved with a single Faddeev equation.
 This calculation requires however some care due to the presence of
open $^3n\to ^2n +n$ channels for some values of the trap parameters
$(V_0,R)$, but will not be developed here.

%

\end{document}